\documentclass[12pt]{iopart}

\usepackage{ifthen}
\usepackage{ifpdf}
\usepackage{color}

\usepackage[section]{placeins}

\ifpdf
\usepackage{graphicx}
\usepackage{epstopdf}
\else
\usepackage{graphicx}
\usepackage{epsfig}
\fi
\graphicspath{{./Figs/}{./}}

\usepackage{latexsym}
\usepackage{amstext}
\usepackage{amssymb}
\usepackage{bm}
\usepackage{wasysym}

\usepackage{hyperref}

\usepackage{fixltx2e}
\usepackage{graphicx}
\usepackage{tikz}
\usepackage{pgfplots}

\DeclareGraphicsExtensions{.pdf,.jpg}

\newcommand{\eexp}[1]{\mathrm{e}^{#1}}

\newcommand{\be}[1]{\begin{eqnarray} \label{e#1}}
\newcommand{\beq}{\begin{eqnarray}}
\newcommand{\eeq}{\end{eqnarray}} 

\newcommand{\hide}[1]{}

\newcommand{\Eq}[1]  {{\textcolor{blue}{Eq.}}(\ref{#1})} 
\newcommand{\hrefl}[1]{\href{#1}{[link]}}

\begin{document}


\title{An Atomtronic Flux Qubit: A ring lattice of Bose-Einstein condensates interrupted by three weak links}

\author{D. Aghamalyan$^{1,2}$, N.T. Nguyen$^2$, F. Auksztol$^2$, K. S. Gan$^2$,  M. Martinez Valado$^2$, P. C. Condylis$^2$, L.-C. Kwek$^{2,3}$, R. Dumke$^{2}$, and  L. Amico$^{2,4,5}$.}

\address{$^1$Institut de Physique de Rennes, UMR 6251 du CNRS and Universit\'e de Rennes 1, 35042 Rennes Cedex, France}
\address{$^2$Centre for Quantum Technologies, National University of Singapore, Singapore 117543}
\address{$^3$ National Institute of Education and Institute of Advanced Studies, Nanyang Technological University, 1 Nanyang Walk, Singapore 637616}
\address{$^4$ CNR-MATIS-IMM $\&$ 
Dipartimento di Fisica e Astronomia,   Via S. Sofia 64, 95127 Catania, Italy}
\address{$^5$ Laboratori Nazionali del Sud, INFN, via S. Sofia 62, 95123 Catania, Italy.}



\begin{abstract}
We study a physical system consisting of a Bose-Einstein condensate confined to a ring shaped lattice potential interrupted by three weak links. The system is assumed to be driven by an effective flux piercing the  ring lattice. 
By employing path integral techniques, we explore the effective quantum dynamics of the system in a pure quantum phase dynamics regime.  Moreover,  the effects of the density's quantum fluctuations are studied  through exact diagonalization analysis of the spectroscopy of the  Bose-Hubbard model. We demonstrate that a clear two-level system emerges by tuning the magnetic flux at degeneracy.
The lattice confinement, platform for the condensate, is realized experimentally  employing a spatial light modulator.    
\end{abstract}


\section{Introduction}




Atomtronics provides for a  new twist in quantum technology\cite{seaman2007atomtronics,pepino2009atomtronic}.  A  defining theme of the field is to confine ultra-cold atoms in laser traps of different intensities and shapes and put them into the motion of coherent flows.  With this approach, quantum devices with a radically new architecture can be conceived; at the same time the scope of quantum simulations could be enlarged considerably\cite{Amico-Boshier}. 

Much interest has been devoted to  Bose-Einstein condensates confined to ring-shaped  potentials that, in a way, form the most elementary Atomtronics circuit\cite{Amico_LG, NIST1,NIST2}. Ring-shaped condensates interrupted by weak links are  Atomtronics Quantum Interference Devices (AQUIDs), in analogy with the SQUIDs conceived in mesoscopic superconductivity\cite{NIST1,PhysRevLett.106.130401,NIST2,BoshirPRL2013,PhysRevLett.113.045305}.  Similar to SQUIDs, AQUIDs enclose  a great potential  both for basic science and technology,  with control of  noise, and reduced decoherence.       They have been suggested, in particular, to provide a physical realization of qubits.  Indeed, rf-AQUID,  a ring-shaped condensate interrupted by a single weak-link, were demonstrated  to be governed by an effective qubit dynamics\cite{hallwood2006macroscopic, solenov2010metastable,Amico2014,Aghamalyan15}. The two states of the qubit are the symmetric and anti-symmetric combinations of the clockwise and anti-clockwise flow-states\cite{leggett1980macroscopic}.  In \cite{Amico2014,Aghamalyan15}, the rf-AQUID is realized by a bosonic fluid with an additional lattice confinement along the ring-shaped potential. The resulting device is the cold atoms analogue of the many-Josephson-junction fluxonium\cite{fluxonium}. Indeed, assuming that the bosons occupy only the lowest Bloch band,  the presence of the lattice helps to set the superfluid current to  the most advantageous regime for the device's operation. For instance, because of the one-dimensional dynamics, the vortex formation rate along the flow is negligible, which should in turn yield a favorable scaling of the qubit gap with the bosonic density\cite{Amico2014} (see also formula (\ref{gap})). Moreover, the lattice provides a natural way to implement a  spatially localized  weak-link (on the scale of the lattice spacing). Finally, it provides an easy way 'infrastructure'  to realize interacting ring--ring architectures\cite{couple_rings}.

Although the physics of the two-level  system can be traced in the ground state of the system\cite{Aghamalyan15,Rey07}, a notoriously challenging  problem (both for SQUIDs and AQUIDs)  is to  address the qubit and the superposition state  experimentally.  For the AQUIDs,  the fingerprint of the effective qubit dynamics would be the onset of Rabi oscillations in the atomic density of the flow-states \cite{Amico2014,Aghamalyan15}.  Besides technical/experimental  difficulties, a possible bottleneck arise from the narrow range of parameters required in order that the system provides a well defined two-level system with appreciable persistent currents\cite{Mooij1999,Feigelman2000}. 

In this paper, we adapt the logic applied in the context of solid state Josephson junctions \cite{Mooij1999,Feigelman2000} to a specific cold atoms setup: We study a Bose-Einstein ring condensate  with three weak links. As for the case of our  rf-AQUID   \cite{Amico2014,Aghamalyan15}, our implementation relies on the lattice potential confinement along the ring.
By a combination of analytic and numerical techniques, we demonstrate that the system can sustain a two-level effective dynamics. We find a set of parameters which considerably enlarge the regime in which the qubit dynamics arise.
We also experimentally realize a trapping potential of a ring-shaped optical lattice with a $\sim 20 \mu m$ diameter using a spatial light modulator. 

The paper is outlined as follows. In the section \ref{models}, we introduce the Hamiltonians describing the microscopic dynamics of the system and review the low energy properties of the rf-AQUID realized with a ring lattice of condensates.  In the section \ref{quantum_dynamics} we derive the path integral effective quantum dynamics of the ring condensate trapped in a ring lattice with three weak links. In the section  \ref{qubit}, we study the system through exact diagonalization. In the section  \ref{platform} we will describe the experiment we carried out to realize the platform for the system. The section \ref{conclu} is devoted to the conclusions.

\section{Models}
\label{models}

{We consider  $N$ Bosons in an $M$~site ring described by the Bose-Hubbard Model. The Hamiltonian reads}
\be{1}
\mathcal{H}_{\text{BHH}} \ \ = \ \ \sum_{i=1}^{M} \left[
\frac{U}{2} \bm{n}_{i}(\bm{n}_{i}-1) 
- {t_i} \left(\eexp{i\Omega} \bm{a}_{i{+}1}^{\dag} \bm{a}_{i} + \text{h.c.} \right)
\right]~.
\label{BH}
\eeq
where $\bm{a}_i \, (\bm{a}^{\dagger}_i)$ are bosonic annihilation (creation) operators on the $i$th site
and $n_{i}= \bm{a}_{i}^{\dag}\bm{a}_{i}$ is the corresponding number operator.
Periodic boundaries are imposed, meaning that $\bm{a}_{M} \equiv \bm{a}_0$.
The parameter $U$ takes into account the finite scattering length for the atomic 
two-body collisions on the same site. The  hopping parameters are constant $t_j=t$ 
except in the three weak-links lattice sites $i_0,i_1,i_2$ where they are  $t_{i_0}=t^{\prime}, t_{i_1}=t_{i_2}=t''$.   
The ring is pierced by an artificial (dimensionless) magnetic flux $\Omega$,
which can be experimentally induced for neutral atoms as a Coriolis flux by rotating the lattice
at constant velocity~\cite{fetter,wright}, 
or as a synthetic gauge flux by imparting a geometric phase
directly to the atoms via suitably designed laser fields~\cite{berry,synth1,dalibard}.
The presence of the flux $\Omega$ in \Eq{e1} has been taken into account through 
the Peierls substitution: $t_i \rightarrow e^{-i\Omega} t_i$. 
{The Hamiltonian~(\ref{e1}) is manifestly periodic in $\Omega$ with period $2\pi$; in addition  it enjoys  the symmetry $\Omega \leftrightarrow -\Omega$. In the absence of the weak-link, the system is also rotationally invariant and therefore the particle-particle interaction energy does not depend on $\Omega$. The many-body ground-state energy, as a function of $\Omega$, is therefore given by a set of parabolas each corresponding to a well defined angular momentum state, shifted with respect to each other by a Galilean transformation and intersecting at the frustration points $\Omega_{n} = (2n + 1)\pi$~\cite{leggett,loss}. 
The presence of the weak-link breaks the axial rotational symmetry and couples different angular momenta states, thus lifting the degeneracy at $\Omega_{n}$. This feature sets the qubit operating point\cite{Amico2014,Aghamalyan15}.}  

It is worth noting that  the interaction $U$ and the weak-link strength induce competing physical effects: the weak-link sets an  healing  length in the density as a 
further spatial scale; the interaction  tends to smooth out the healing length effect. As a result, strong interactions tends to renormalize the weak link 
energy scale\cite{Aghamalyan15,Minguzzi}.

%
%
%
In the limit of a large number of bosons in each well $\bar{n}=N/M$, $a_i\sim \sqrt{ \bar{n}}e^{i\phi_i} $,  and the Bose-Hubbard hamiltonian (BHH) \Eq{e1} can be mapped to the Quantum-Phase model 
employed to describe Josephson junction arrays\cite{Amico2000,fazio2001quantum}:
%
%
\beq
\mathcal{H} \ \ = \ \  \sum_{i=1}^{M} 
\left[
\frac{U}{2} \bm{n}_{i}^2 
- J_i  \cos\left(\bm{\phi}_{i{+}1}{-}\bm{\phi}_{i} - {\Omega}\right) 
\right]   
\label{H_qp}
\eeq
where  $\left [\bm{n}_i , \bm{\phi}_l\right ]=i \hbar\delta_{il}$ are canonically conjugate number-phase variables and $J_i\sim \bar{n} t_i$ are the Josephson tunneling amplitudes.

\paragraph{The rf-AQUID qubit.} In this case, a single weak link occurs along the ring lattice $t''=t$.
The presence of the weak link induces a slow/fast  separation of the effective (imaginary time) dynamics: the dynamical variables relative to the weak link are slow compared 
to the 'bulk' ones, playing the role of an effective bath (nonetheless, we assume that the ring system is perfectly isolated by the environment). Applying the harmonic approximation to the fast dynamics and integrating it out,  the  effective dynamics 
of the AQUID  is governed by \cite{Amico2014}
%
\be{5}
\mathcal{H}_{\text{eff}} \ \ = 
 \mathcal{H}_{\text{syst}}+
 \mathcal{H}_{\text{bath}} +  \mathcal{H}_{\text{syst-bath}}
\label{H_eff}
\eeq
The slow dynamics is controlled by
\begin{equation}
\mathcal{H}_{\text{syst}}=U  \bm{n}^2 
+ E_L \bm{\varphi}^2
- E_J \ \cos(\bm{\varphi}-\Omega)  
\label{jj}
\end{equation} 
where $ \varphi$ is the phase slip across the weak link,  with  $E_L=J /M$, and $E_J=J^{\prime}$.  
For $\delta \doteq {E_J}/{E_L} \ge 1 $, $\mathcal{H}_{\text{syst}}$ describes a particle in a double well potential with the two-minima-well separated from the other
 features of the potential.  The two parameters, U and $t'/t$, allow control of the two level system. The  two local minima of the double well are degenerate for  ${\Omega=\pi}$.
The minima correspond to the clock-wise and anti-clockwise currents in the AQUID.
Because of the quantum tunneling between the two minima of the double well, the two states of the system (qubit)
are formed by symmetric and antisymmetric combinations of the two circulating states. 
%
%
%
%
 %
 The WKB level splitting is  
\begin{equation}
\Delta \simeq \frac{2 \sqrt{UE_J}}{\pi}\sqrt{ (1-\frac{1}{\delta})} e^{-12\sqrt{ E_J/U} (1-1/\delta)^{3/2} }\;.
\label{gap}
\end{equation}
From this formula we can see that the limit of a weak barrier and intermediate to strong interactions form the most favourable regime to obtain a finite gap between the two energy levels of the double level potential\cite{Aghamalyan15}.
Incidentally, we comment that the bath  Hamiltonian in  Eqs.(\ref{H_eff}), (\ref{jj}), is similar to the one
describing the dissipative dynamics of a  single Josephson junction in the framework of the Caldeira-Leggett model\cite{Caldeira-Leggett}. As long as the ring has finite size, however, 
there are a finite number of discrete modes and no real dissipation occurs\cite{rastelli}.
In  the limit $N \rightarrow \infty$, a proper Caldeira-Leggett model is recovered.
%
%
%
%
%
In agreement to the arguments reported above, the qubit dynamics encoded in the AQUID is less and less addressable by increasing the size of the ring
\cite{Amico2014,Aghamalyan15}.    

Finally, we observe that the condition ${E_J}/{E_L} \ge 1 $ imposes specific constraints on the physical parameters that may be difficult to fulfill. For example, because of the mesoscopic nature of the persistent currents, $M$ cannot be too large. On the other hand, small values of $J'/J$ may block persistent currents (see \cite{Aghamalyan15} for a thorough analysis).    We shall see that the proposed architecture with three-weak-links realize a two-level system with a wider range of exploitable parameters.

\section{Effective quantum dynamics}
\label{quantum_dynamics}
In this section, we derive the  low energy effective phase dynamics of the system.
To this end, we elaborate on the imaginary-time $\tau$ path integral of the partition function of the model Eq.(\ref{BH}) in the limit of large fluctuations of the number of bosons at each site\cite{rastelli,Amico2014}.  We first perform a local gauge transformation
$a_l\rightarrow  a_l e^{i l \Omega}$ eliminating the contribution of the magnetic field everywhere except in one of  the weak link sites\cite{twisted}). Here, we refer to the  Quantum-Phase Hamiltonian (\ref{H_qp}):
\begin{eqnarray}
\hspace*{-2.5cm} H_{QP}=\sum_{i=1\atop i \neq i_0,i_1,i_2}^{M} \left [ {{U\over2}}  n_i^2 -J \cos\left ( \phi_{i+1}-\phi_{i}\right )\right ]   + \sum_{\alpha=0,1,2} \left [  {{U\over2}}  n_{i_\alpha}^2-J_\alpha \cos\left ( \phi_{i_\alpha+1}-\phi_{i_\alpha}-\Omega_\alpha \right )\right ]
\end{eqnarray}
where the weak links are placed at the sites $i_0,i_1,i_2$ with  $\Omega_0=\Omega$, $\Omega_1=\Omega_2=0$.

The partition function of the model Eq.(\ref{H_qp}) is
\begin{equation}
Z=Tr \left ( e^{-\beta H_{BH}} \right )  \propto \int D[\{ \phi_i \}]  e^{-S[\{ \phi_i \}]}
\end{equation}
where the action is
\begin{eqnarray}
S[\{\phi_i\}]=\int_0^\beta d\tau\sum_{i=1\atop i \neq i_0,i_1,i_2}^{M}  &\left [ {\frac{1}{U/2}} (\dot{\phi}_{i})^2-J \cos\left ( \phi_{i+1}-\phi_{i}\right )\right ]   + \nonumber \\
 &\sum_{\alpha=0,1,2}  \left [ {\frac{1}{U/2}} (\dot{\phi}_{i_\alpha})^2-J_\alpha \cos\left ( \theta_\alpha-\Omega_\alpha \right ) \right ]
\end{eqnarray}
where $\theta_\alpha\doteq  \phi_{i_\alpha+1}-\phi_{i_\alpha} $.  We assume that the weak links are sufficiently spaced   to 
make the nearest neighbour phase differences in between them (fast variables) small. This implies that substantial phase  slips occur at the weak links with the constraint $\theta_0+\theta_1-\theta_2=0 \; mod(2\pi)$. The goal, now,  is to integrate out the phase variables in the 'bulk'.   With our assumption,  the harmonic approximation can be applied to the bulk phases
\begin{eqnarray}
\sum_{i=0\atop i \neq i_0,i_1,i_2}^{M} \cos\left ( \phi_{i+1}-\phi_{i}\right ) \simeq \left (M-3 \right )-\sum_{i=0\atop i \neq i_0,i_1,i_2}^{M}    {{\left ( \phi_{i+1}-\phi_{i}\right )^2}\over{2}}  \;.
\end{eqnarray}
The cases  $i=\left \{i_0\pm1,i_1\pm1, i_2\pm1\right \}$ involve phase variables coupled to $\theta_\alpha$. 
Let us define: $\phi_{i_\alpha}\doteq -{{\theta_\alpha}\over{2}} +\varphi_\alpha$ and  $\phi_{i_\alpha+1}\doteq {{\theta_\alpha}\over{2}} +\varphi_\alpha$. Then
\begin{eqnarray}
\sum_{i=1\atop i \neq i_0,i_1,i_2}^{M}    &{\left ( \phi_{i+1}-\phi_{i}\right )^2}=\sum_{\alpha=0,1,2}  [{{1}\over{2}}(\theta_\alpha)^2 + (\phi_{i_\alpha-1}-\phi_{i_\alpha+2}) \theta_\alpha+ \nonumber\\
&\hspace*{-2cm}+ (\varphi_\alpha-\phi_{i_\alpha-1})^2+(\varphi_\alpha-\phi_{i_\alpha+2})^2]+
\sum_{ i \neq \left \{ i_0,i_1,i_2,i_0\pm1,i_1\pm1,i_2\pm1 \right \}}  (\phi_i-\phi_{i+1})^2= \nonumber \\
&\sum_{\alpha=0,1,2}  [{{1}\over{2}}(\theta_\alpha)^2 + (\psi_{j_\alpha-1}-\psi_{j_\alpha+1}) \theta_\alpha]+ 
\sum_{j=1}^{M-3}    {\left ( \psi_{j+1}-\psi_{j}\right )^2}
\end{eqnarray}
where $\{ \phi_1,\dots \phi_{i_0-1},\varphi_{0},  \phi_{i_0+2}\dots \phi_{i_1-1}, \varphi_{1}, \phi_{i_1+2}\dots \phi_{i_2-1}, \varphi_{2},\phi_{i_2+2}\dots \phi_{M}\}\equiv \{ \psi_j \, ,\, j=1,\dots, M-3 \} $.

The effective action, $S[\{\phi_i\}]$, can be split into two terms
$
S[\{\phi_i\}]=S_{1}[\{\theta_\alpha\}] +S_{2}[\{\psi_{j}\}, \{\theta_\alpha\}]
$
with
\begin{eqnarray}
&& \hspace*{-2cm}S_{1}[\theta_\alpha]=\int d\tau \left [ \sum_\alpha {\frac{1}{U/2}} (\dot{\theta_\alpha})^2+ {{J}\over{4}} \theta_\alpha^2-J_\alpha \cos\left (\theta_\alpha -\Omega_\alpha \right )\right ]  \label{S_theta} \\
&& \hspace*{-2cm}S_{2}[\{\psi_{j}\}, \{ \theta_\alpha\} ]=\int d\tau \left \{ \sum_{j=0}^{M-3} \left [ {\frac{1}{U/2}} (\dot{\psi}_{j})^2 +{{J}\over{2}}   \left ( \psi_{j+1}-\psi_{j}\right )^2 \right ]+ {{J}\over{2}}    \theta_\alpha (\psi_{j_\alpha-1}-\psi_{j_\alpha+1})  \right \} \nonumber
\label{S_coupling}
\end{eqnarray}

For our device, $J_0=J'$, $J_1=J_2=J''$, and we   fix $i_0\equiv 1$  and $i_1,i_2$  equidistant from $i_0$. In this  way, the system enjoys a mirror symmetry around $i_0$:
$\psi_j={{1}\over{\sqrt{M-3}}} \sum_{k=1}^{(M-4)/2}\psi_k \sin[2\pi kj/(M-3)]$ (the cases corresponding to general configurations of the weak links will be studied elsewhere): 
\begin{equation}
S_{2}[\{\psi_{j}\}, \theta] =\int d\tau {{1}\over{U/2}}\sum_k (\dot{\psi}_{k})^2 +\omega_k^2 \psi_{k}^2+ {{U}\over{2}} J \theta_\alpha  {\zeta_\alpha}_k \psi_{k}
\end{equation}
where $ \omega_k= \sqrt{J U/2 \left [ 1-\cos{ \left ( {{2\pi k}\over{M-3}} \right )} \right ] } $ and 
$${\zeta_\alpha}_k= {{1}\over{2}}\left (\sin[2\pi k(j_\alpha-1)/(M-3)]-\sin[2\pi k(j_\alpha+1)/(M-3)]\right )/\sqrt{M-3} \; .$$ 
The integral in $\{\psi_{k} \} $ involves  the interaction with the harmonic modes describing the phase slips away from the  weak links. Such integration is standard (see f.i. \cite{path_integral_book} for a general textbook and  \cite{rastelli,rastelli1} for recent applications particularly relevant for the present problem)
leads to a nonlocal kernel in the imaginary time: $\int d\tau d\tau' \theta(\tau) G_\alpha(\tau-\tau')\theta(\tau')$. The explicit form of $G(\tau-\tau') $ is obtained by expanding  $\{\psi_{k} \} $ and $\theta_\alpha$ in Matsubara frequencies $\omega_l=2\pi l/\beta$: $\displaystyle{ \psi_k (\tau)=\sum_{l=0}^\infty \psi_{kl} e^{i\omega_l \tau} }$, $ \displaystyle{\theta_\alpha(\tau)=\sum_{l=0}^\infty \theta_{\alpha l} e^{i\omega_l \tau} }$. Therefore
\begin{equation}
\int D[\psi_{k}] e^{-\int d \tau S_{2}}\propto \exp{\left (-\beta {{UJ^2}\over{2}} \sum_\alpha \sum_{l=0}^{\infty} \tilde{Y_\alpha}(\omega_l) |{\theta_\alpha}_l |^2  \right )}
\end{equation}
with $\tilde{Y_\alpha}(\omega_l)=\sum_{k=1}^{(M-4)/2} {{{\zeta_\alpha}_k^2}\over{\omega_k^2+\omega_l^2}}$. The $\tau=\tau'$ term is extracted by summing and subtracting $\tilde{Y_\alpha}(\omega_l=0)$. The effective action reads
\begin{equation}
\hspace*{-2cm}S_{eff}=\sum_{\alpha=0,1,2} \int_0^\beta d \tau \left [ {\frac {1}{ U}}  \dot{\theta_\alpha}^2 + V(\theta_\alpha) \right ]
- \frac{J}{ U} \int d \tau d \tau' \theta_\alpha(\tau)G_\alpha(\tau-\tau') \theta_\alpha(\tau')
\label{effective_single}
\end{equation}
where
\begin{eqnarray}
 \hspace*{-1cm}V(\theta_\alpha) \doteq  J    c_\alpha \theta_\alpha^2 -{{J'}\over{3}} \cos (\theta_1-\theta_2-\Omega)-{{J''}\over{3}}(\cos\theta_1+\cos\theta_2) 
\label{potential}
\end{eqnarray}
with  $\displaystyle{c_\alpha=  {{1}\over{2}} \left ( {{1}\over{2}}-U J \sum_{k=1}^{{M-4}\over{2}} {{{\zeta_\alpha}_k^2}\over{\omega_k^2}} \right )}$. The interaction between the fast and the slow modes is described by the kernel
\begin{equation}
 \hspace*{-1cm} G_\alpha(\tau)=\sum_{l=0}^\infty  \sum_{k=1}^{{M-4}\over{2}} {\frac{\omega_l^2 {\zeta_\alpha}_k^2}{\omega_k^2+\omega_l^2}} e^{i\omega_l \tau}\; .
\end{equation}
\begin{figure*}[!t]
\centering
  \includegraphics[width=0.6\textwidth]{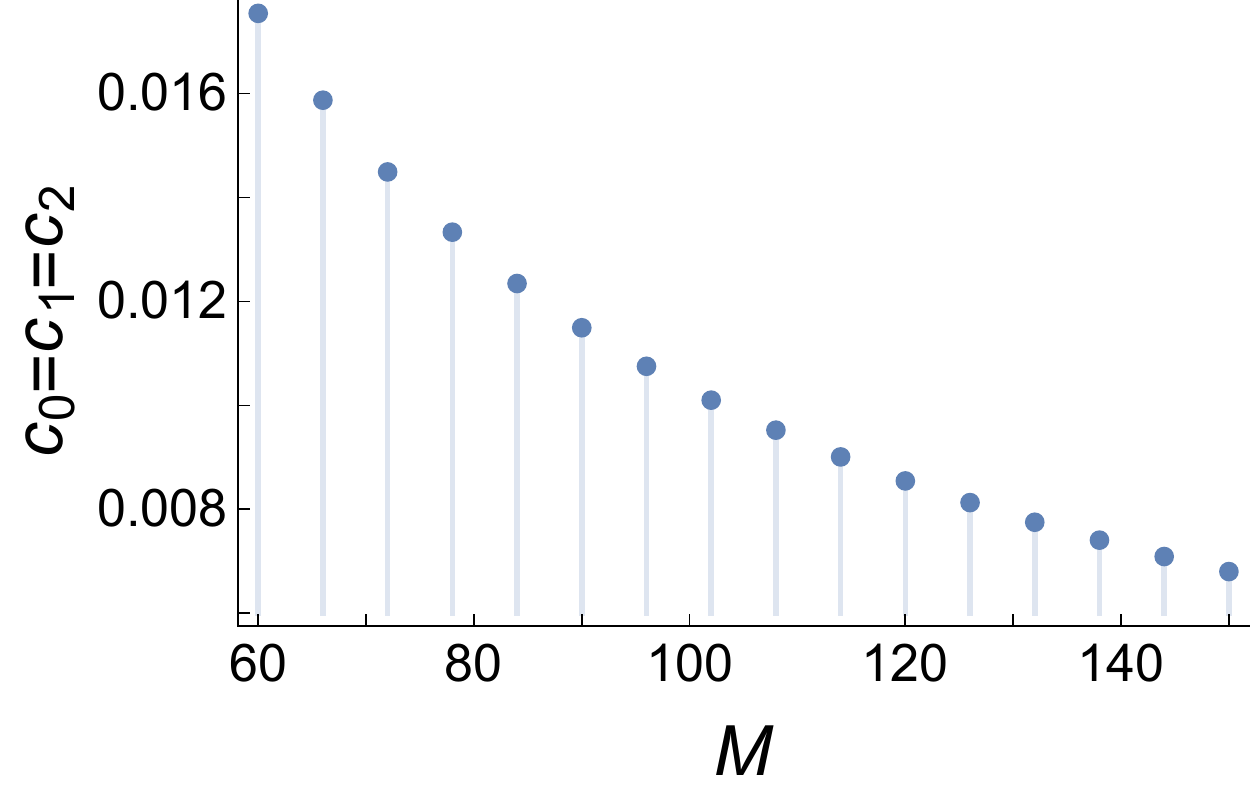}
  \caption{Scaling of the coefficients $c_\alpha$ in (\ref{potential}) with the number of sites in the lattice ring for equidistant weak links ($M$ is even and multiple of $3$)} 
  \label{scaling}
\end{figure*}

\begin{figure*}[!t]
\centering
  \includegraphics[width=0.55\textwidth]{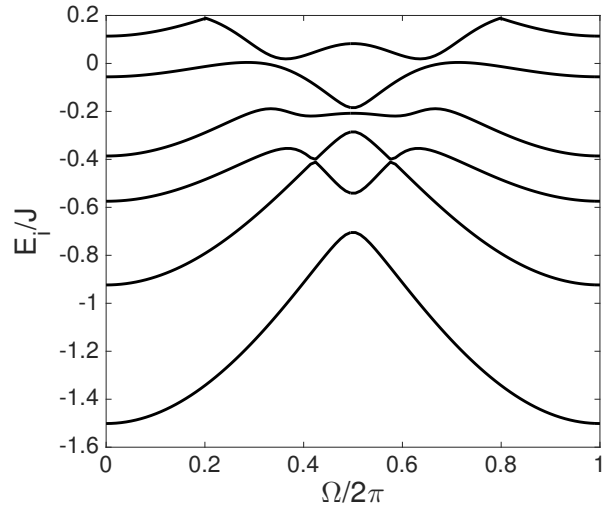}

  \caption{Six first energy levels of the reduced system given by the effective potential Eq. (\ref{potential})   as a function of the dimensionless external flux $\Omega$. Here ${{J'}}=0.7J$, ${{J''}}=0.8J$, $U=0.5J$, and $\theta_1=-\theta_2$. } 
  \label{SpectrumQF}
\end{figure*}

We observe that $V(\theta_\alpha) $ defines the  effective dynamics of the superconducting Josephson junctions flux qubits\cite{Mooij1999, Feigelman2000}, but perturbed by the $\theta^2$ terms; by numerical inspection, we see that  the corresponding coefficients are small in units of J, and decreases by increasing $M$ (see Fig.~\ref{scaling}). Moreover, on Fig.\ref{SpectrumQF} we introduce the numerical result for the spectrum of the quantum particle which moves in the potential given by Eq. (\ref{potential}) under the additional assumption that $\theta^2$ terms do not contribute. From this figure we clearly see that near the frustration point $\Omega=\pi$ two lowest energy levels are well separated from each other and from higher excitations, which means that effective dynamics of the system defines a qubit. 
%
\section{Spectroscopy of the three weak junctions AQUID}
\label{qubit}

In this section we study the low-lying spectrum of the Bose-Hubbard model~(\ref{BH})  in the low filling regime with the help of the exact diagonalization (ED) method, which is particularly well suited for the small systems with small  total number of particles\cite{Feshke, Zhang}. This kind of analysis has been implemented previously for the AQUID devices with one weak link\cite{Aghamalyan15} and the technical details regarding the implementation of the ED method can be found in the Appendix material of the aforementioned article.
We stress that in contrast with the quantum phase model here we take into account the effect of the number fluctuations and hence of the amplitude of the superfluid order parameter. The presence of  finite weak links, breaks the axial rotational symmetry
and couples different angular momentum states, thus lifting the degeneracy 
at $\Omega=\pi$ by an amount $\Delta E_1$. For all the cases studied with the ED method in this manuscript,   the impurities are not assumed to be equidistant (in contrast to Sect.\ref{quantum_dynamics}): $t_2=t^{\prime}$ and $t_5=t_8=t''$.
Provided other excitations are energetically far enough from the two lowest energy states,
this will identify the two-level system defining the desired qubit and its working point. In order to analyze whether the qubit can perform well we study two quantities which play a central role for this manuscript: the qubit energy gap $\Delta E_1=E_1-E_0$ and the so-called qubit quality factor $\Delta E_1/\Delta E_2$, where $E_0$,  $E_1$ and  $E_2$ are the ground state, first excited state, and the second excited state, respectively, of the many-body Bose-Hubbard Hamiltonian given by the Eq.(\ref{BH}). 

For a functional qubit it is required that in the vicinity of the frustration point ($\Omega=\pi$), the qubit has a finite energy gap $\Delta E_1$, and that quality factor is smaller than $0.5$ (when $\Delta E_1/\Delta E_2 = 0.5$, then the many-body Hamiltonian has equidistant energy levels, which is normally an unfavourable case for defining a qubit since the second excited level can be easily populated. When $\Delta E_1/\Delta E_2 > 0.5$ the two lowest energy levels are quite close to the other energy levels, which increases the decoherence of the qubit by populating higher energy levels).

The average value of the persistent current  in the mesoscopic ring optical lattice can be defined via the following thermodynamic relation \cite{Bloch,Marco}:
\begin{equation}
I(\Omega)= - \frac{1}{2 \pi \hslash}\frac{\partial E_i(\Omega)}{\partial \Omega}
\end{equation}
The geometrical meaning  of the persistent  current is given by the slope of the tangent line to the graph of the i-th energy level $E_i$, when it is plotted as a function of the external artificial flux $\Omega$. Persistent currents are mesoscopic phenomenon, and as it was evidenced in Refs.\cite{Minguzzi}, persistent current amplitudes for the ring potential with a delta-barrier scales as a power law with the system size. The power law scaling was evidenced as well for the energy gap $\Delta E_1=E_1-E_0$   in Ref.\cite{Aghamalyan15}. In order to have a good energy gap and sufficient amplitude of the persistent current we restrict our discussion in the current manuscript only to the mesoscopic ring lattices(M$\sim$ 10). 

\begin{figure*}[!t]
\centering
  \includegraphics[width=0.4\textwidth]{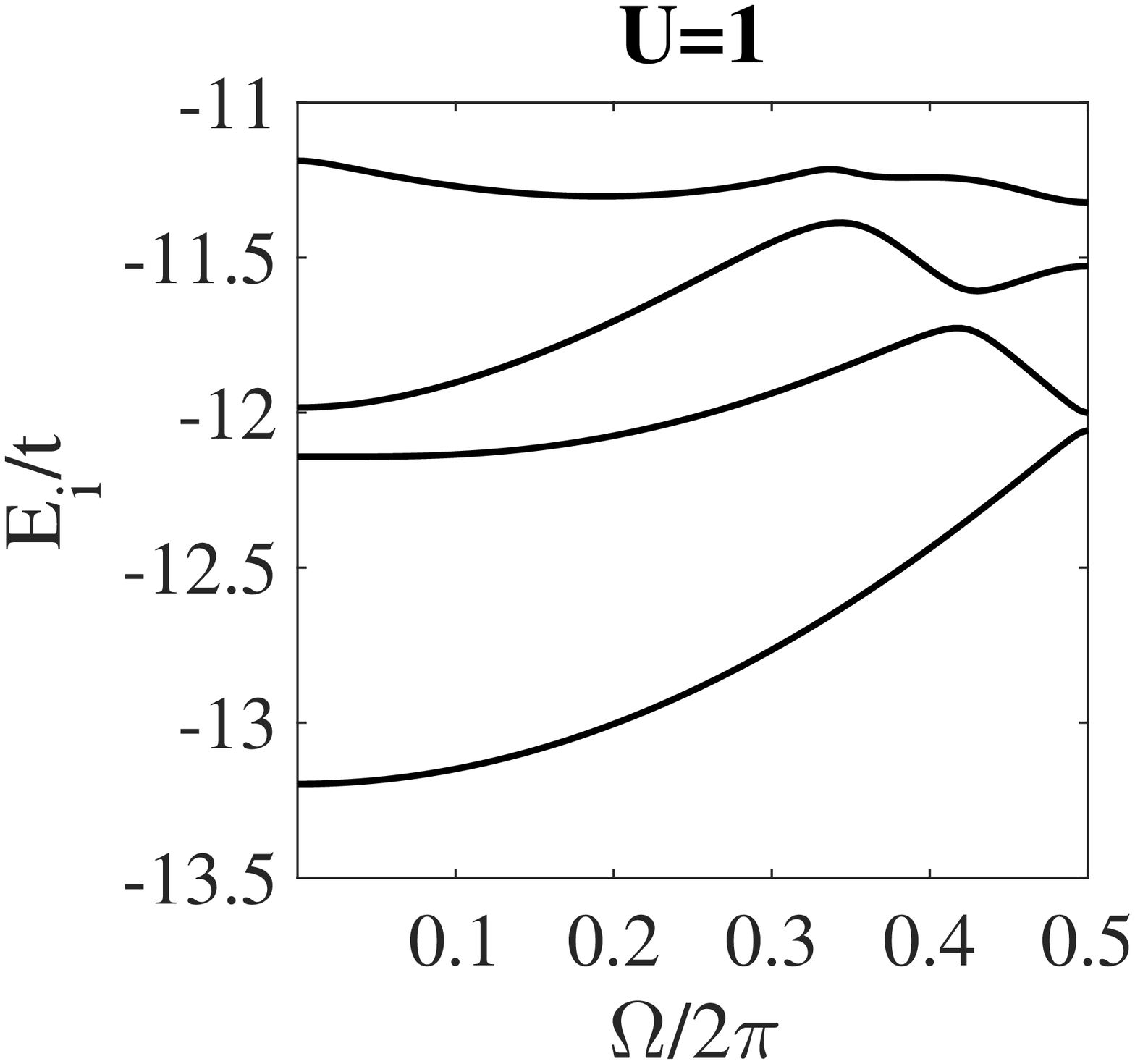}
  \includegraphics[width=0.4\textwidth]{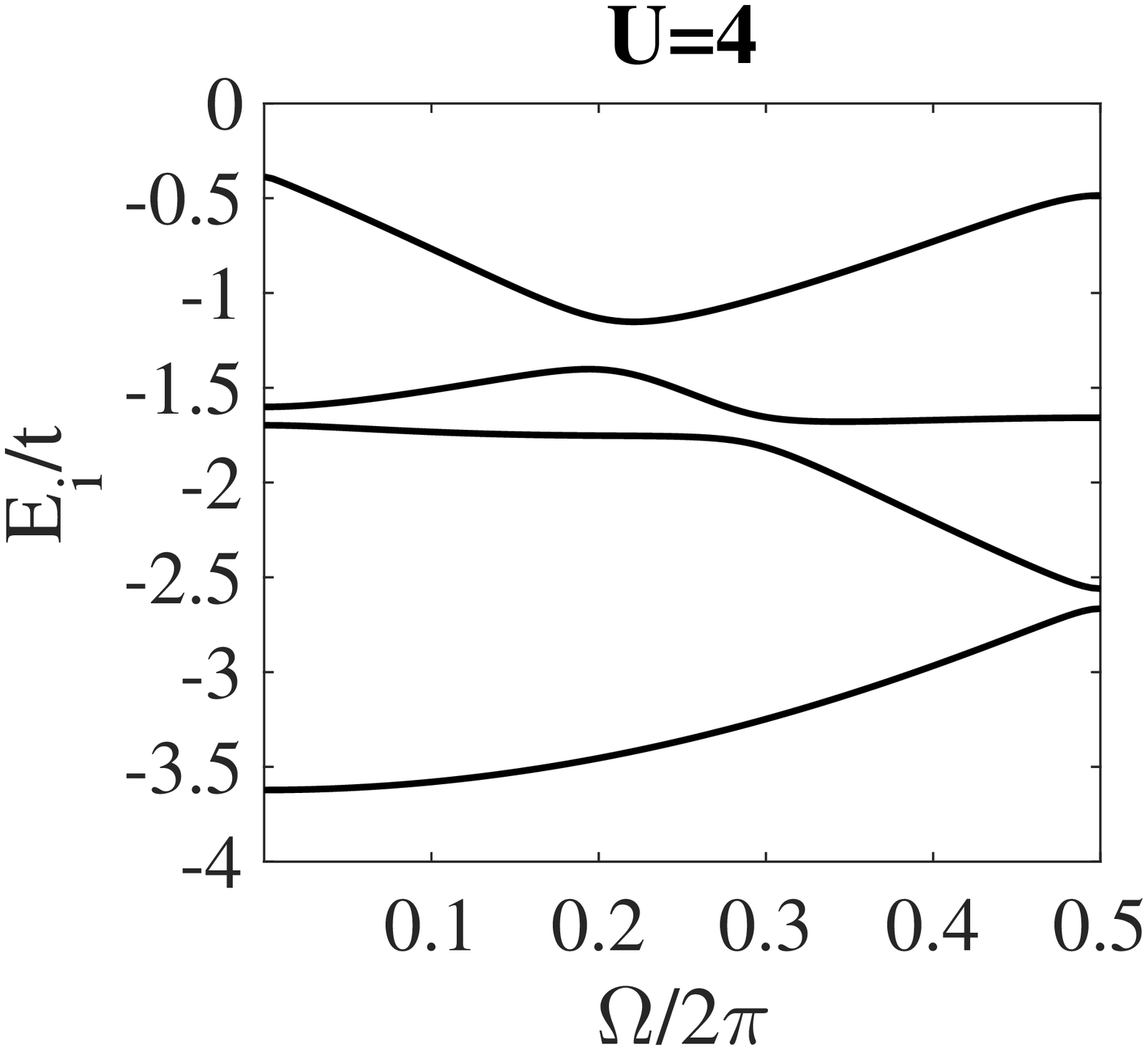} \\
   \includegraphics[width=0.4\textwidth]{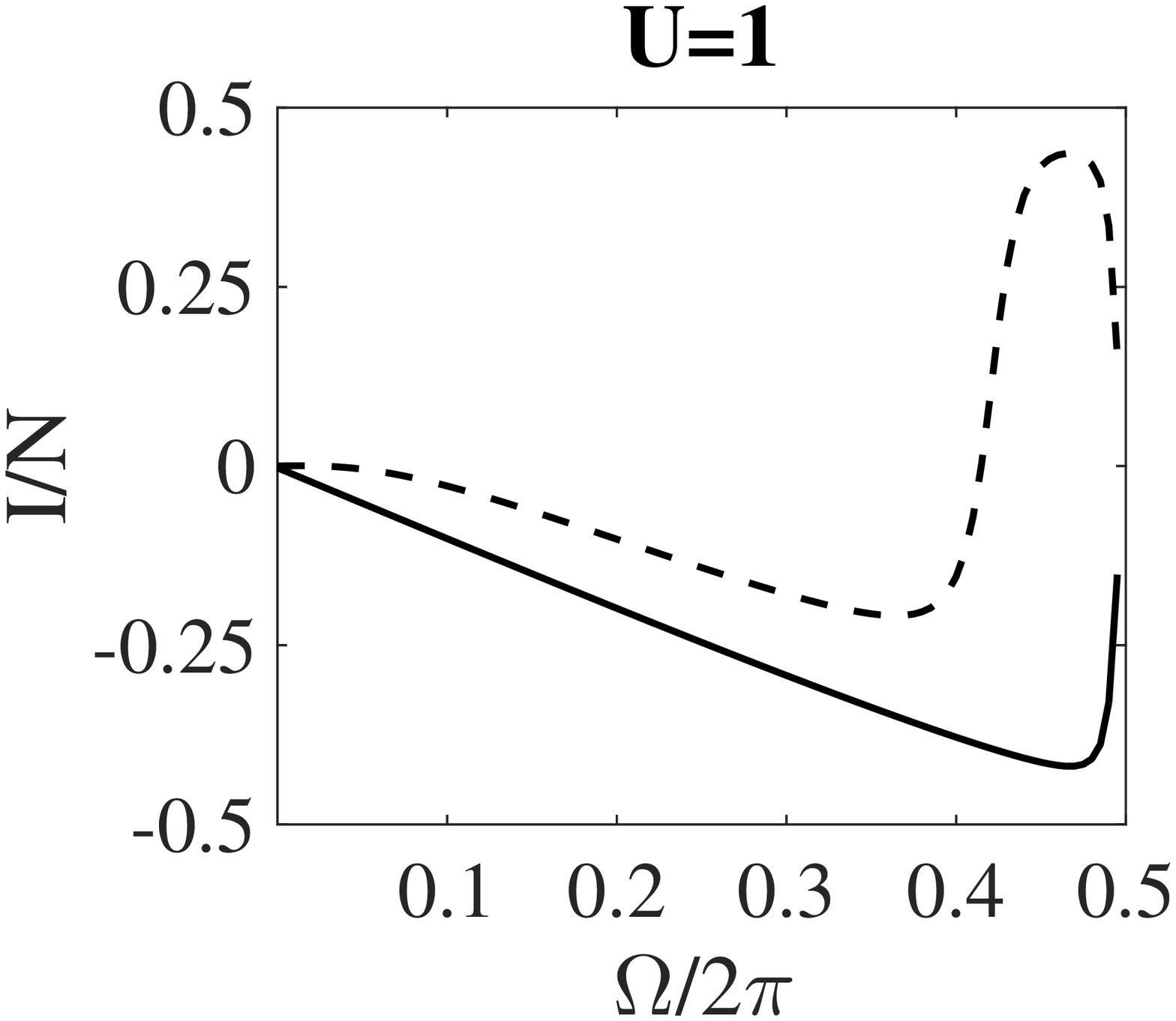}
  \includegraphics[width=0.4\textwidth]{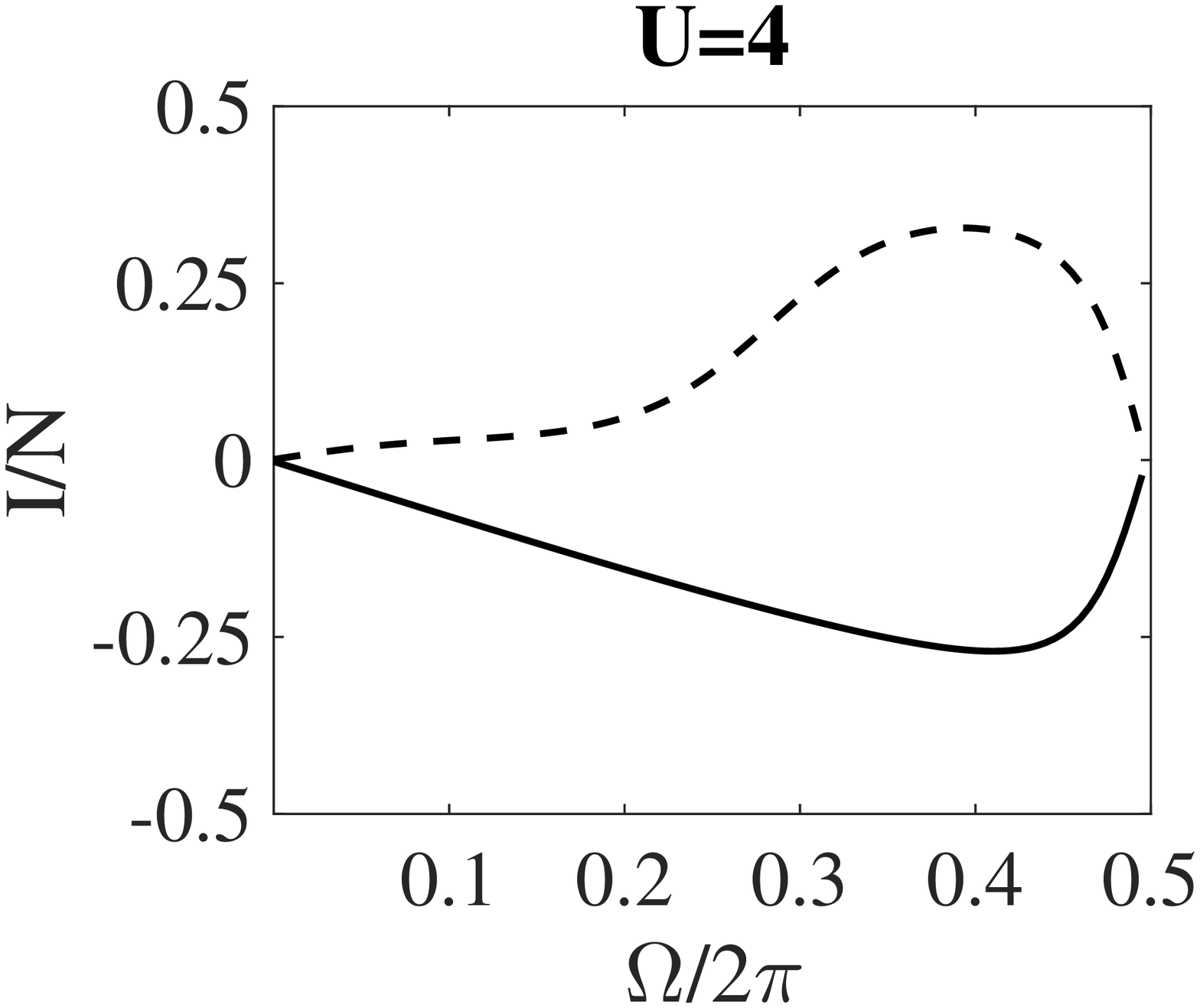} \\
  
  \caption{Upper Panel: low-energy spectrum of the Bose-Hubbard model for various values of the interaction. The four lowest energy levels as function of $\Omega/2 \pi$. Lower panel: Persistent currents in the ground state(solid line) and the first excited stat(dashed line) as a function of the dimensionless flux. Here persistent current defined in such a way that its amplitude is 1. Here M=8, N=10, $t^{\prime}=0.5t$ and  $t''=0.8t$; $t$ is the energy unit.}
  \label{Spectr}
\end{figure*}

The low energy spectrum of the Hamiltonian given by Eq.(\ref{BH}) as a function of the dimensionless flux $\Omega/2 \pi$  is demonstrated on the upper panels of  Fig.\ref{Spectr}. It is clearly seen that near the degeneracy point   $\Omega=\pi$ the qubit has a finite energy gap($\Delta E_1/t \sim 0.05(U=1); 0.25(U=4)$)
and the ground and the first excited states are well separated($\Delta E_1/\Delta E_2 \sim 0.1(U=1); 0.23(U=4)$) from the higher energy levels. 

As demonstrated in the previous study \cite{Aghamalyan15}, it is possible to obtain a good qubit  only  in the range of interactions varying from strong to the infinitely strong. However, as it can be seen from the left panel of the Fig. \ref{Spectr}, in the case of mild interaction $U=1$ it is possible to define a good qubit in the vicinity of the frustration point by introducing additional weak links in the system. It is interesting to point out that in the vicinity of the frustration point, mesoscopic currents in the ground state and the first excited state have the opposite signs as can be seen from the lower panels of Fig.\ref{Spectr}, which means that the two quantum states of the qubit are given by the clockwise and anti-clockwise circulating currents. Moreover, as can be seen from the same panels of Fig.\ref{Spectr}, for the same strengths of the weak links, the amplitude of the persistent current is bigger in the limit of the weakly interacting system, which in turn improves the qubit state readout.

The dependence on the strength of interaction  of the qubit energy gap and the quality factor (at the degeneracy point) is shown in Fig.\ref{EG,QF_U} for the different values of the two equal weak links $t''$. First of all, as can be seen from both panels, the quantities of interest display non-monotonous behaviour as a function of interaction, which has previously been observed for the case of the rf-AQUID\cite{Nunnenkamp08, Aghamalyan15} for the ring lattice systems. This kind of non-monotonous dependence occurs because of the non-trivial screening of the barrier which was thoroughly analyzed in Refs. \cite{Minguzzi}. It is also easy to notice that qubit energy gap is also a non-monotonous function of $t''$ for the interactions ranging  from mild to strong, and is monotonously increasing with $t''$ in the limit of very weak interactions. The non-monotonous behaviour is due to the overlap of healing lengths in the small sized system which gives rise to the non-trivial density distribution along the ring, demonstrated in Section \ref{Densityprofiles}. It is also clearly seen from these figures that if one fixes the interaction strength it is possible to adjust both the qubit energy gap and the quality factor by changing the $t''$. This provides at our disposal an extra parameter for defining a good qubit and for realizing single and two-qubit gates (the authors of the current article will address the issue regarding the qubit gates in a future publication). In the limit of very weak interactions it is not possible to define a good qubit since $\Delta E_1/\Delta E_2 = 0.5$, the quality factor significantly improves with growing interaction strength and it is well below $0.5$ for the interactions ranging from mild to strong. After reaching its minimum value the energy gap increases with the strength of the interaction, however it is important to note that for large values of the interactions the amplitude of the persistent current is smaller than for the intermediate values of the interaction  (a similar trend was noticed in rf-AQUID  Refs.\cite{Minguzzi}). We conclude that a functional qubit can be defined for interactions ranging from mild to strong by adjusting the strength of the two additional weak links $t''$, however we stress that the highest currents in the system are obtained for the mild interaction strengths. In the Section \ref{EG,QF_N} we will demonstrate that our approach improves the scaling of the energy gap with the filling factor in the limit of mild interactions($U=1$).

\begin{figure}[!t]
\centering
  \includegraphics[width=0.4\textwidth]{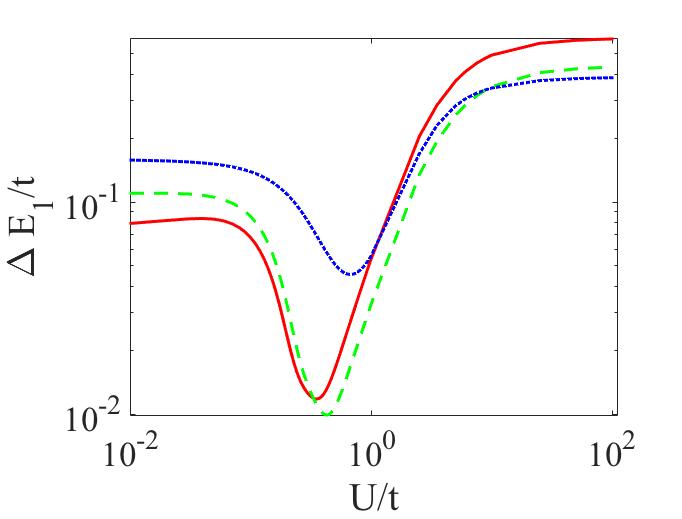}
   \includegraphics[width=0.4\textwidth]{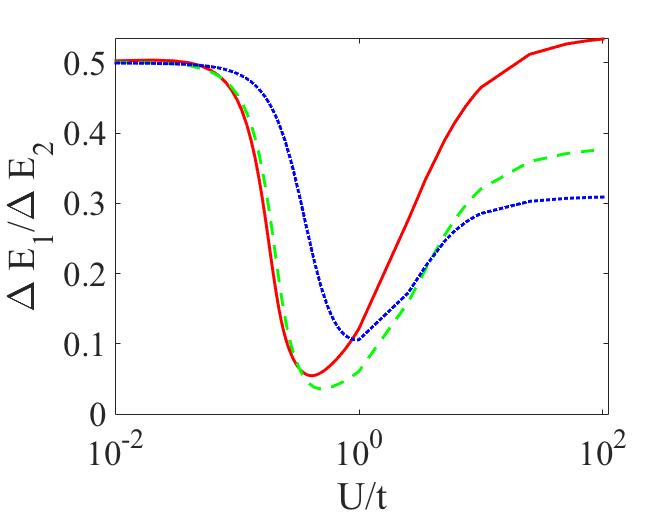}
   \caption{A qubit energy gap $\Delta E_1$ (left panel, which is plotted in the log-log scale) in units of $t$ and the qubit quality factor (right panel, which is plotted in the semi-log scale) as a function of the interaction $U/t$.
   Here we consider $M=8$ , $\Omega=\pi$, $N=10$ and $t^{\prime}=0.5t$. In each plot various curves represent $t''=0.4t$ (red solid lines), $0.6t$ (green dashed lines), and $0.8t$ (blue dotted lines). }
  \label{EG,QF_U}
\end{figure}

\subsection{Density profiles}
\label{Densityprofiles}
In this section we focus on the density distribution of cold atoms along the ring when the system is in the ground state and close to the qubit working point, $\Omega=\pi$.

\begin{figure*}[!t]
\centering
  \includegraphics[width=0.35\textwidth]{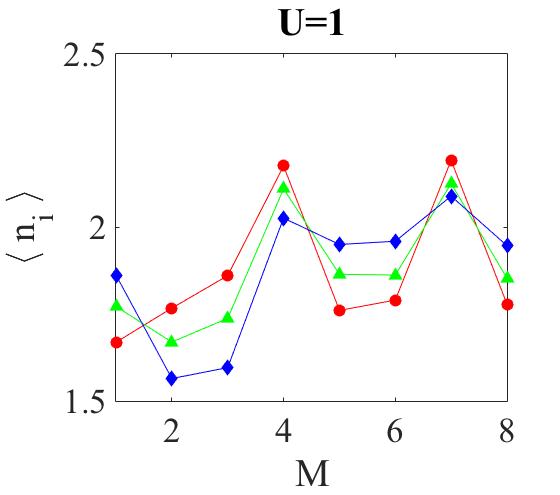}
  \includegraphics[width=0.35\textwidth]{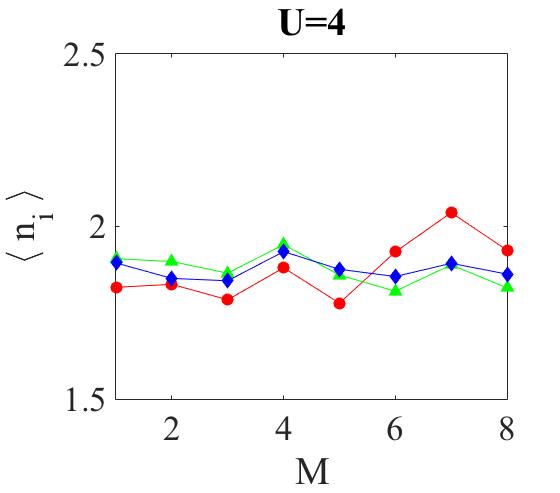}
  \includegraphics[width=0.35\textwidth]{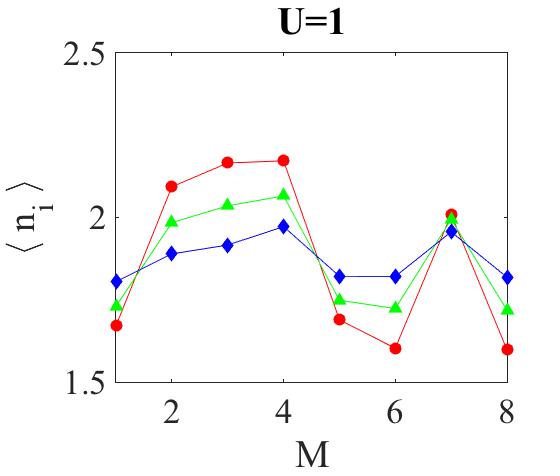}
  \includegraphics[width=0.35\textwidth]{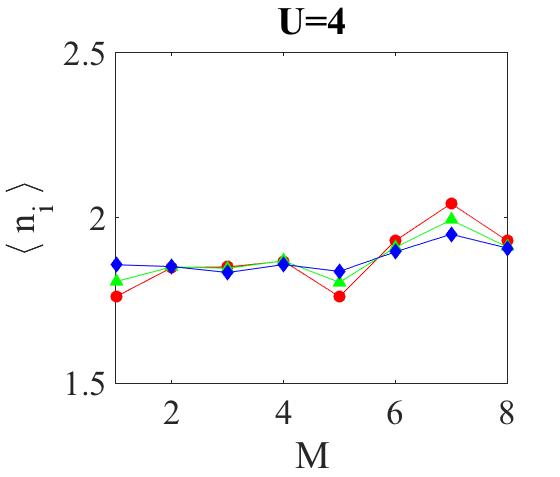}
  \caption{Density profiles $\langle n_j \rangle$
    at $\Omega = \pi$, along the ring with $M=8$ sites, $N=15$ particles and $t^{\prime}=0.5$ (Top), $t^{\prime}=0.9$ (Bottom) for different interaction regimes.
    In each plot various curves represent $t''=0.4t$ (red circles), $0.6t$ (green triangles), and $0.8t$  (blue diamonds). $t$ is the energy unit.}
  \label{dens}
\end{figure*}

An evident effect of the barrier is a reduction of the particle density 
in its  vicinity; depending on the ring size,
the whole density profile along the ring may well be affected.
The interplay between the interaction strength $U$ and the strength of the weak links
implies different behaviors, as exemplified in Fig.~\ref{dens}
for the mesoscopic ring. The depth of the density depression in the case of one impurity increases monotonously 
with the barrier strength\cite{Aghamalyan15,Minguzzi}, while its width decreases with increasing $U$  
since the density can be reduced at the cost of multi-occupancy of the other sites; which means with increasing interaction strength multiple occupancy and thus the healing length tends to decrease  in order to minimize the ground state energy.
As it is clearly seen from the left panel of Fig. \ref{dens}, the density distribution at the lattice sites 5 and 8 decreases with increasing barrier strength (which is equivalent to the decreasing hopping strength $t''$). However,  at lattice site 2 even if we keep the same barrier strength $t'$  still the density is affected because of the overlap of the healing lengths of the weak links. Moreover, the biggest density depression is achieved when the other two links have the smallest barrier strengths $t''=0.8t$ (which is given by the blue curve). And subsequently, the density depression decreases with increasing $t''$ at the lattice site 2.

As explained previously, with increasing interactions the healing length decreases which is clearly seen from the right panel of Fig. \ref{dens}. However the appearance of Friedel oscillations is apparent especially for the case of strong weak links given by the red curves. This kind of oscillations are a result of the  strong correlations of 1D bosons 
that respond to impurities similarly to fermions since for strong interactions the system undergoes fermionization\cite{Tonks}. We finally observe that, for moderate interaction ($U=1$ in Fig. \ref{dens}), the weak links are effective enough to  determine commensurate effects on a tripartite ring.   For  stronger interaction, instead, the weak links are washed out  with the density distribution along the ring that is nearly uniform  along the ring; in Section \ref{GAP_N} we will  see that such a  feature  implies  commensurate effects and Mott peaks at  values of the filling of the uniform system.

\subsection{Scaling of the qubit energy spectrum with the filling}
\label{EG,QF_N}

\begin{figure}[!t]
\centering
  \includegraphics[width=0.3\textwidth]{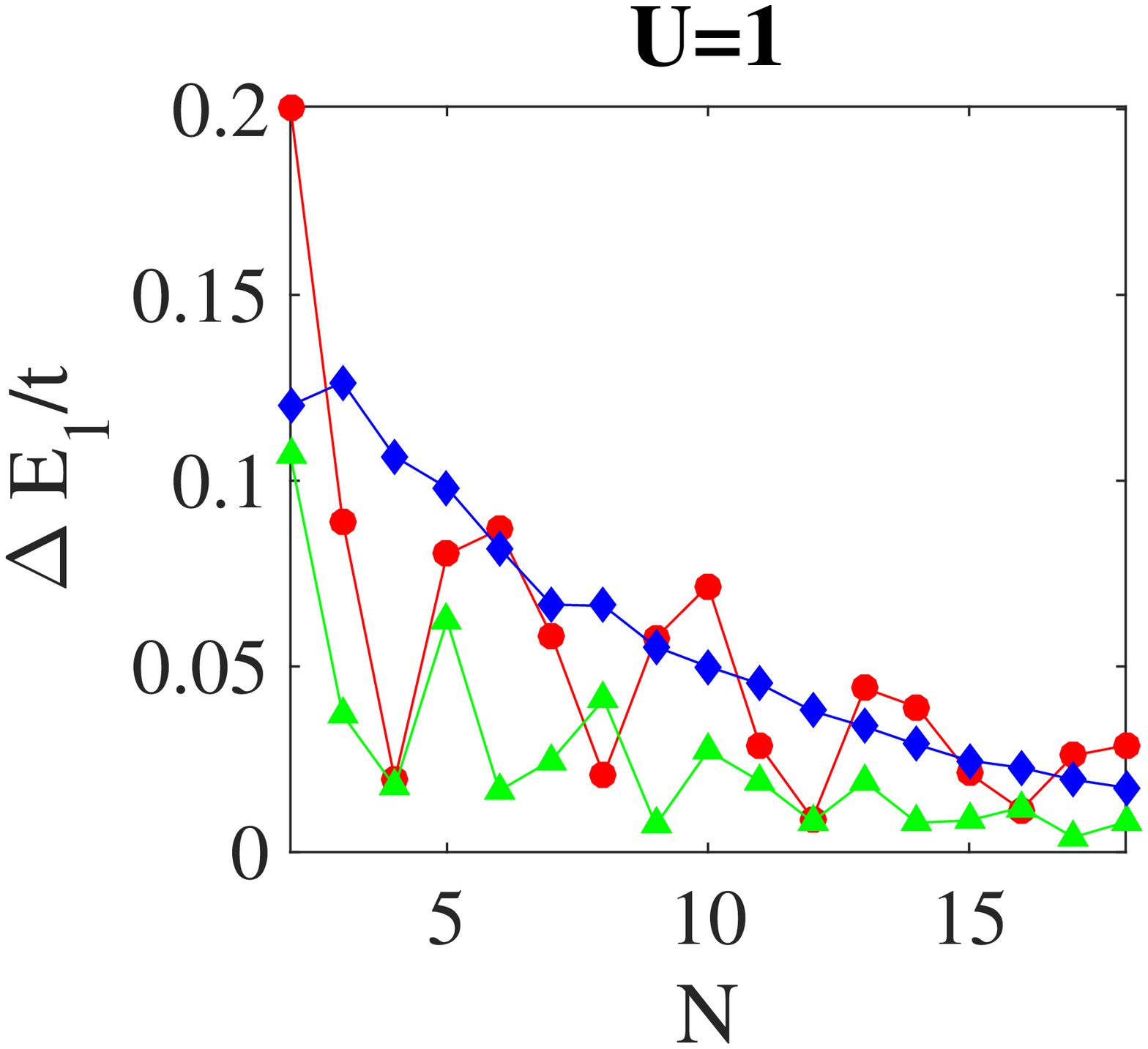}
  \includegraphics[width=0.3\textwidth]{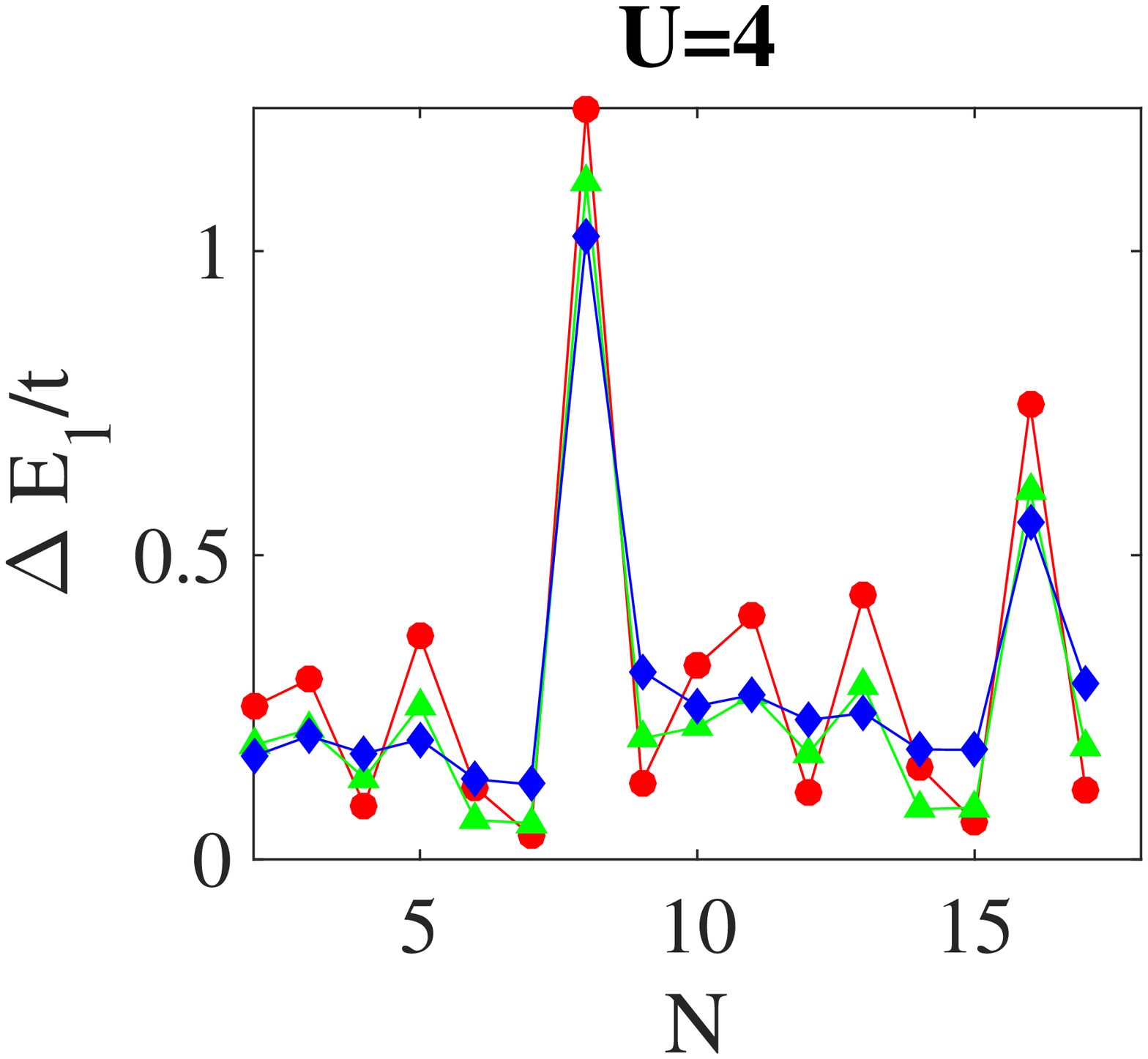}\\
  \includegraphics[width=0.3\textwidth]{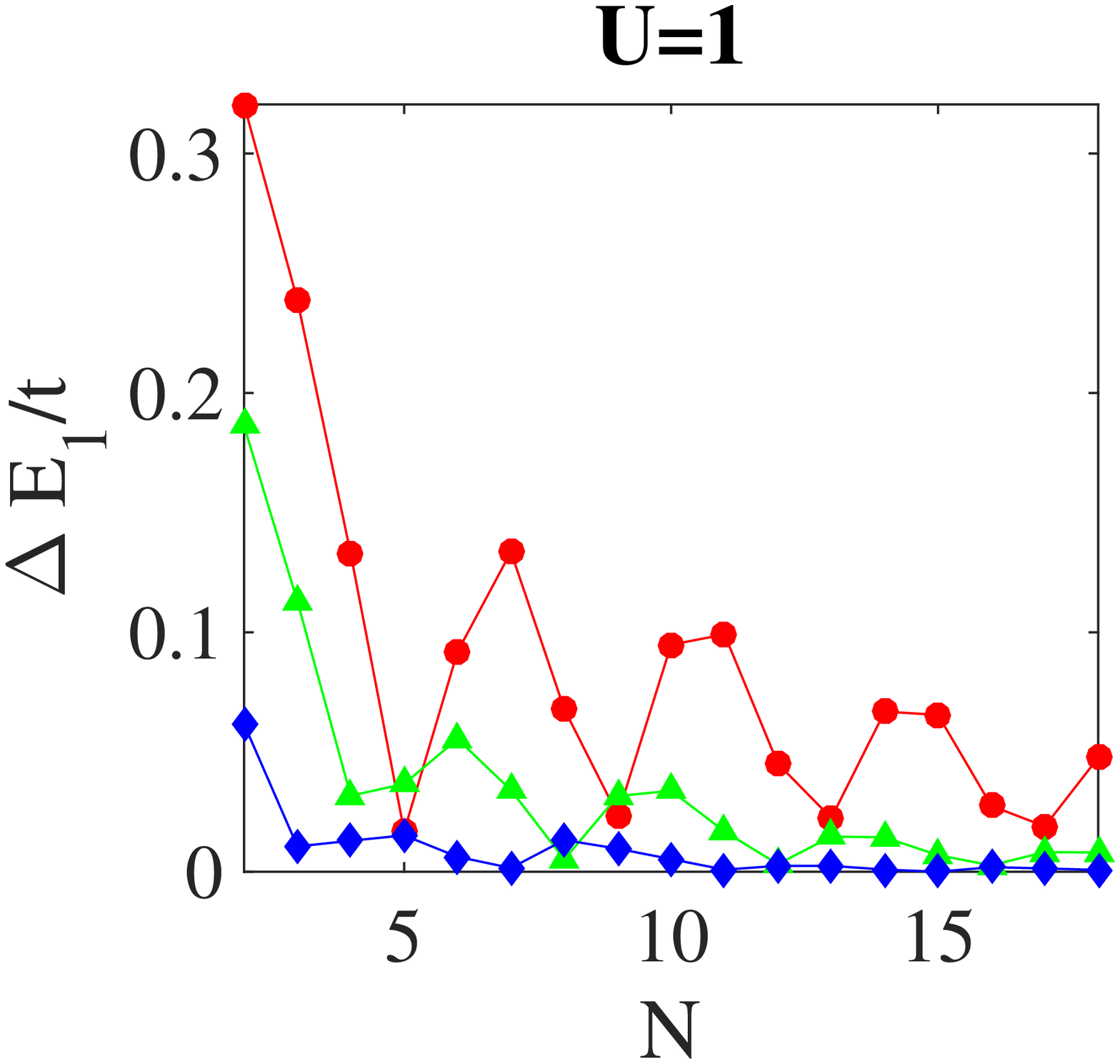}
  \includegraphics[width=0.3\textwidth]{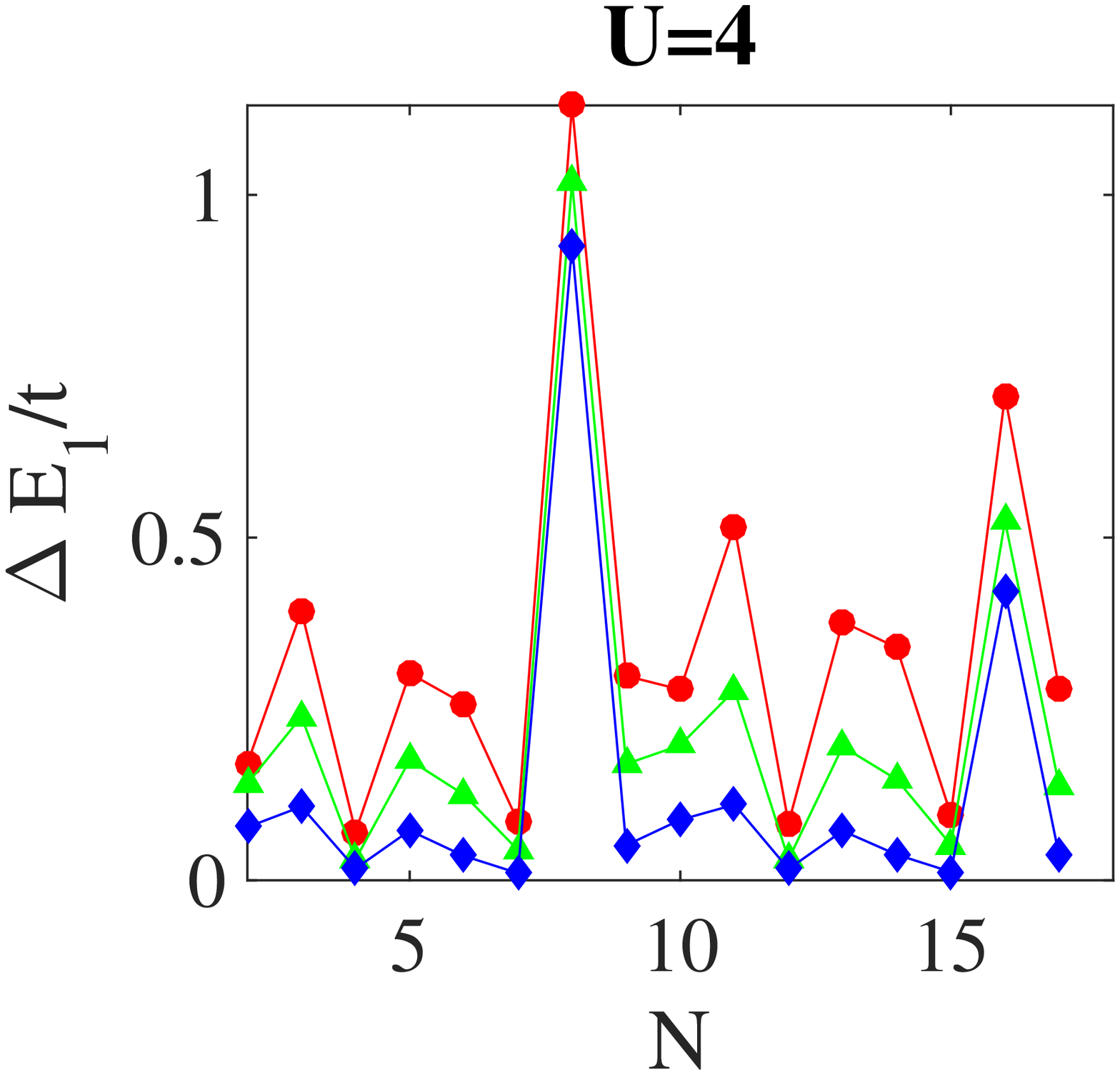}
   \caption{Many-body energy gap $\Delta E_1$ in units of $t$ as a function of the total particle number $N$.
   Here we consider $M=8$ lattice sites and $\Omega=\pi$ and $t^{\prime}=0.5t$ (Top), $t^{\prime}=0.9$ (Bottom). In each panel various curves represent $t''=0.4t$(red circles), $0.6t$ (green triangles), $0.8t$(blue diamonds). $t$ is the energy unit.}
  \label{GAP_N}
\end{figure}

In Fig.~\ref{GAP_N}  we present our results for the qubit energy gap 
 at the frustration point for two different $t'$, $t'=0.5$ (top), and $t'=0.9$ (bottom)
as a function of the filling at fixed system size,
studying its dependence on the strength of the two equal weak links $t''$ for the different interaction strength values.

The top panels of Fig.~\ref{GAP_N} present the data for fixed interaction strengths for $\Delta E_1$
($U/t = 1$ and $U/t=4$, respectively)
with different curves representing $t''=0.4t;0.6t;0.8t$.

At small $U$ (Fig.~\ref{GAP_N} top-left), we observe two kinds of behaviours for the energy gap: monotonously decreasing ($t''=0.8t$) and monotonously decreasing but modulated with an oscillatory behaviour $t''=0.4t;0.6t$. First of all, we remark that compared to the case of the rf-AQUID in \cite{Aghamalyan15} introducing extra impurities in the system improves the scaling of the energy gap with the filling factor, which can be clearly seen for the case given by the blue curve. The reason for the very fast decrease of the gap in the case of the rf-AQUID is the fact that the healing length scales as $\sim1/\sqrt{nU}$, and consequently soon enough the system is not affected by the barrier, which opens the gap at the frustration point. As it was pointed out in Section \ref{qubit} having several weak links in the system modifies significantly the dependence of the healing length from the particle number and the interaction strength due to the effect of the overlap of several healing lengths (this effect is also clearly seen from the density profiles analysed in the previous section). We also notice that the energy gap and quality factor have non-monotonous dependence on $t''$ as it was already evident in section \ref{EG,QF_U}. We conclude, that the values of the weak links $t''=0.6;0.8$ are favourable for defining a good qubit with a finite energy gap ($\Delta E_1 \sim 0.05t $) and good quality factor ($\Delta E_1/\Delta E_2 \sim 0.1$). 
\begin{figure}[!t]
\centering
   \includegraphics[width=0.3\textwidth]{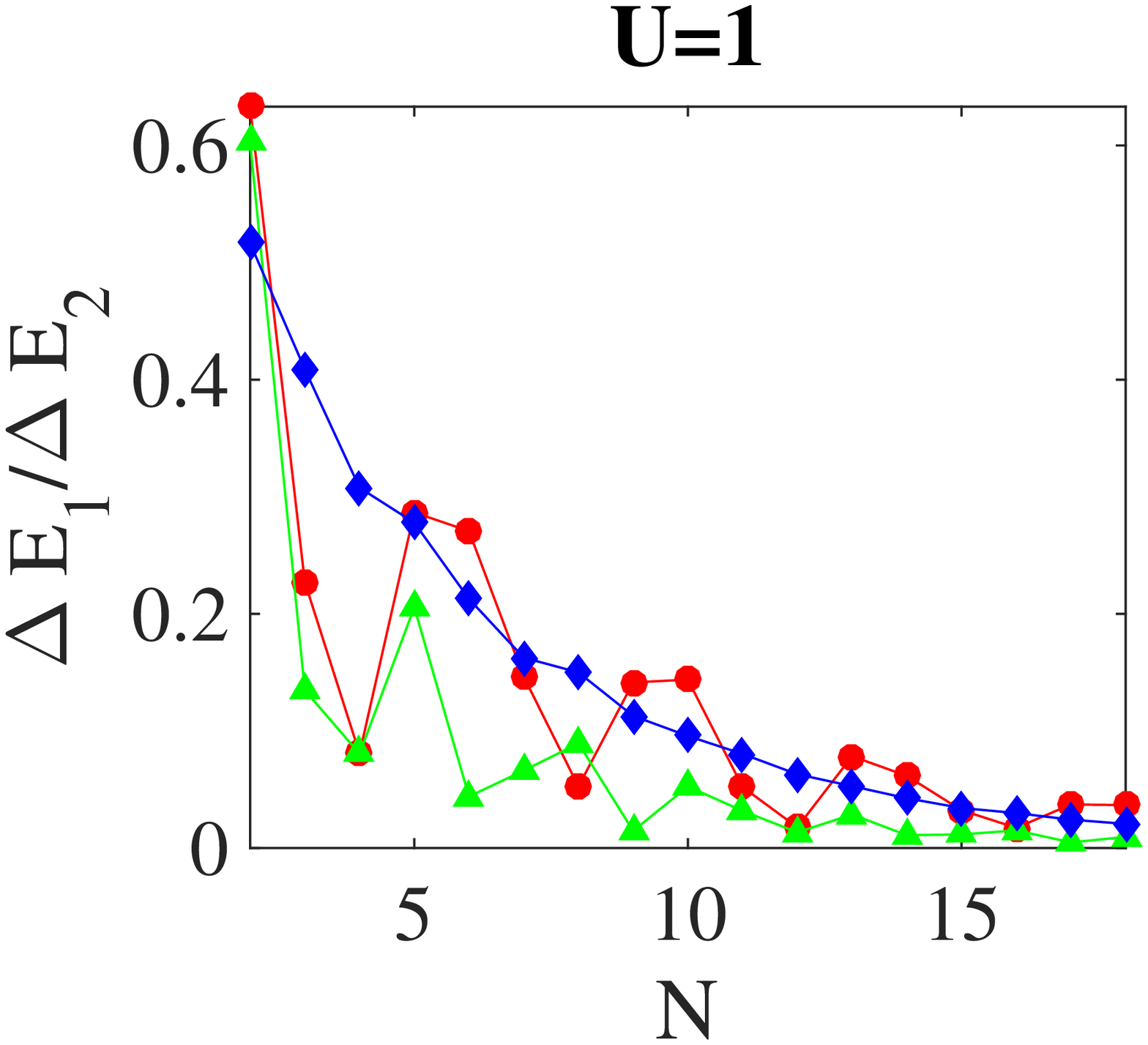}
  \includegraphics[width=0.3\textwidth]{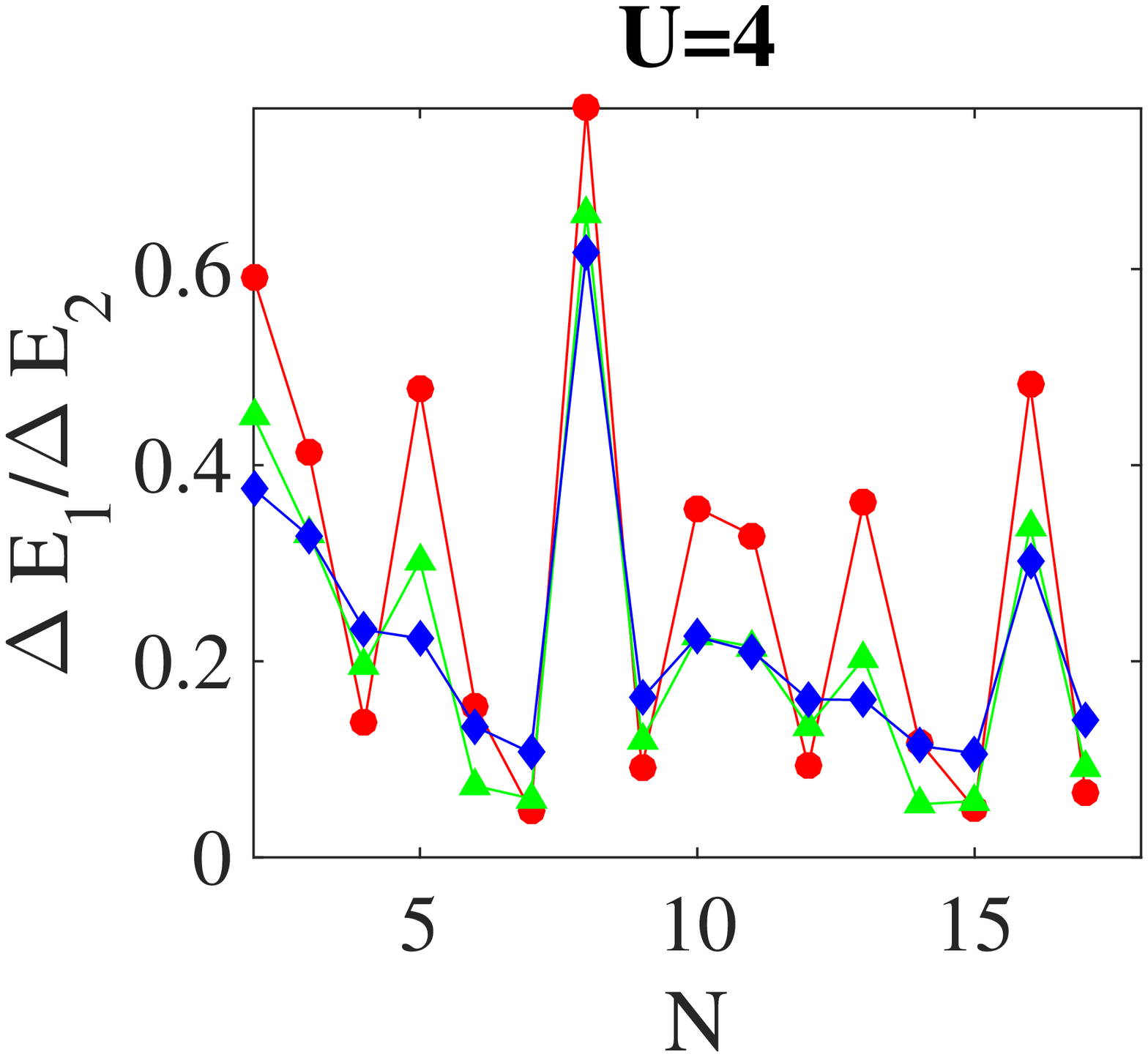}\\
    \includegraphics[width=0.3\textwidth]{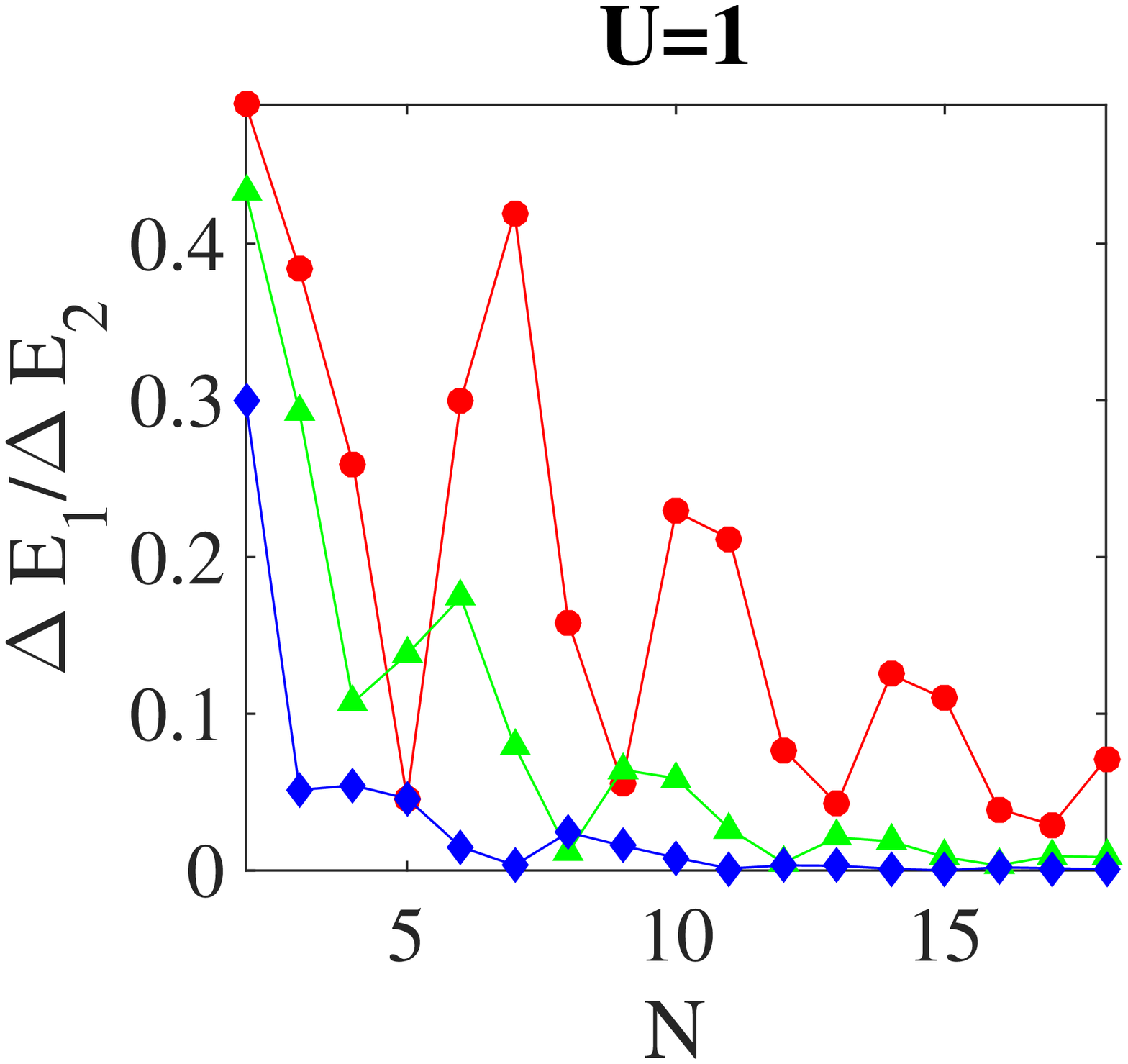}
  \includegraphics[width=0.3\textwidth]{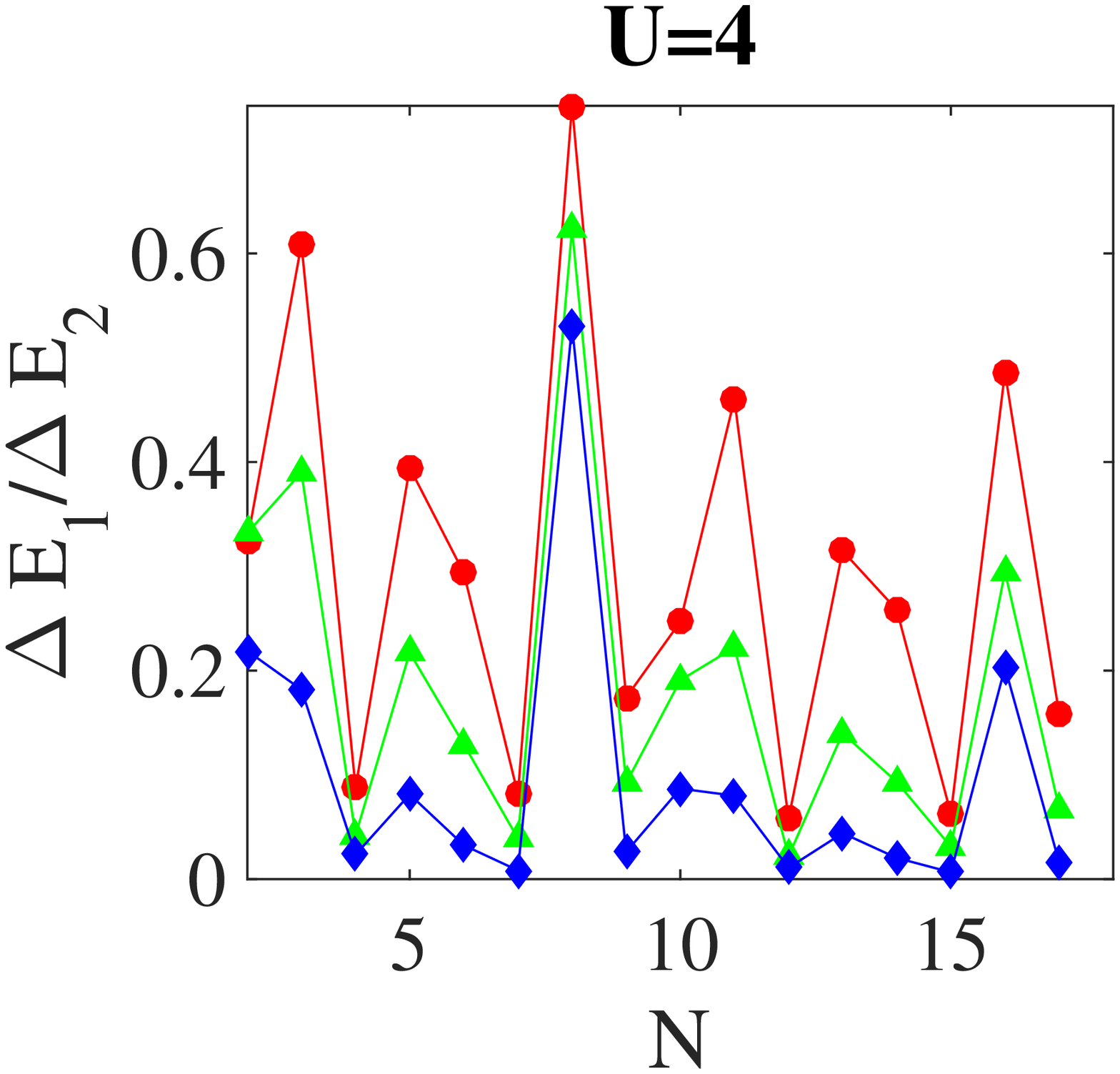}
   \caption{Many-body qubit quality factor as a function of the total particle number $N$.
   Here we consider $M=8$ lattice sites and $\Omega=\pi$ and $t^{\prime}=0.5t$ (Top), $t^{\prime}=0.9$ (Bottom). In each panel various curves represent $t''=0.4t$(red circles), $0.6t$ (green triangles), $0.8t$(blue diamonds). $t$ is the energy unit.}
  \label{QF_N}
\end{figure}

For large values of the interaction strength (Fig.~\ref{GAP_N}) the system effectively enters into the Mott regime. We comment that the superfluid to Mott transition normally occurs in the thermodynamic limit. However, signatures of this transition can be seen even in the mesoscopic system under study. An apparent signature of the Mott phase is the appearance of the Mott peaks for the values of commensurate fillings (which in our case corresponds to the number of particles $N=8;16$).
It is interesting to point out that in the case of the rf-AQUID, for a strong barrier, the Mott peaks appear not exactly at the commensurate filling, which occurs because the strong barrier effectively opens the ring and the system reduces to an $M-1$ lattice site ring. However, in the case of three weak links, the Mott peaks appear exactly at the commensurate filling and this can be understood from the fact that the density distribution in this case is almost homogeneous along the ring (see right panel Fig \ref{dens}). We also conclude from Fig. \ref{GAP_N} that for the values of $t''=0.6t;0.8t$,  a good qubit can be obtained with $\Delta E_1 \sim  0.3 $ and $\Delta E_1/\Delta E_2 \sim 0.3$ for all values of the filling except for commensurate fillings, where the quality factor becomes $0.5$, see Fig. \ref{QF_N}. This is unfavourable for defining a functional qubit.

\section{Experiment} 
\label{platform}
We produce the optical potential using a spacial light modulator (Holey Photonics AG, PLUTO-NIR II), SLM \cite{Amico2014}. A collimated Gaussian beam, of 8 mm diameter, is reflected from the SLM's surface forming an image through a $f=200$ mm lens. The light is then split into the two sides of our system, with 10\% of the light in the ``monitoring" arm, and 90\% into the ``trapping" arm used to create a red-detuned dipole trapping potential for a gas of Rb\textsuperscript{87} atoms. A Ti:Saph laser (Coherent MBR-110)  produces a 1W, 828 nm beam, which is spacially filtered and collimated, before reflection on the SLM. To produce the trapping potential the SLM's kinoform is imaged through a 4\textit{f} lens system reducing the beam size to 3 mm diameter and focused through a 50X microscope objective with a 4 mm focal distance and a numerical aperture of NA = 0.42 (Mitutoyo 50X NIR M-Plan APO). The monitoring arm of the system creates an image of the potential through a 10X infinity-corrected microscope objective focused on a CCD camera (PointGrey FL3-GE-13S2M-C). The CCD camera views, therefore, an enlarged image of the optical potential.

The SLM is comprised of a liquid-crystal-on-silicon display of 1920 x 1080 8 $\mu$m pixels, with 8-bit phase values from 0 to 2$\pi$, covering a 15.36 x 8.64 mm region with a filling factor of 87\%. The individual pixels create a phase-shift pattern (kinoform) on the incoming light, which can be used to create arbitrary 2D optical potentials on the image-plane of a Fourier transforming lens. To construct the kinoform applied to the SLM we used an improved version of the Mixed-Region-Amplitude-Freedom (MRAF) algorithm \cite{pasienski2008high, SLM} with angular spectrum propagator. A region outside of the desired pattern is used to collect unwanted light contributions which result from the iterative MRAF algorithm. The MRAF algorithm iteratively finds a solution to minimise the error, within the region of interest, between the computationally produced image (the Fourier transform of the kinoform) and the desired result. However, when this kinoform is applied to the SLM there arise additional errors due to imperfections in the SLM and aberrations in the optics.

\begin{figure}[htb]
    \centerline{\includegraphics[width=0.75\textwidth]{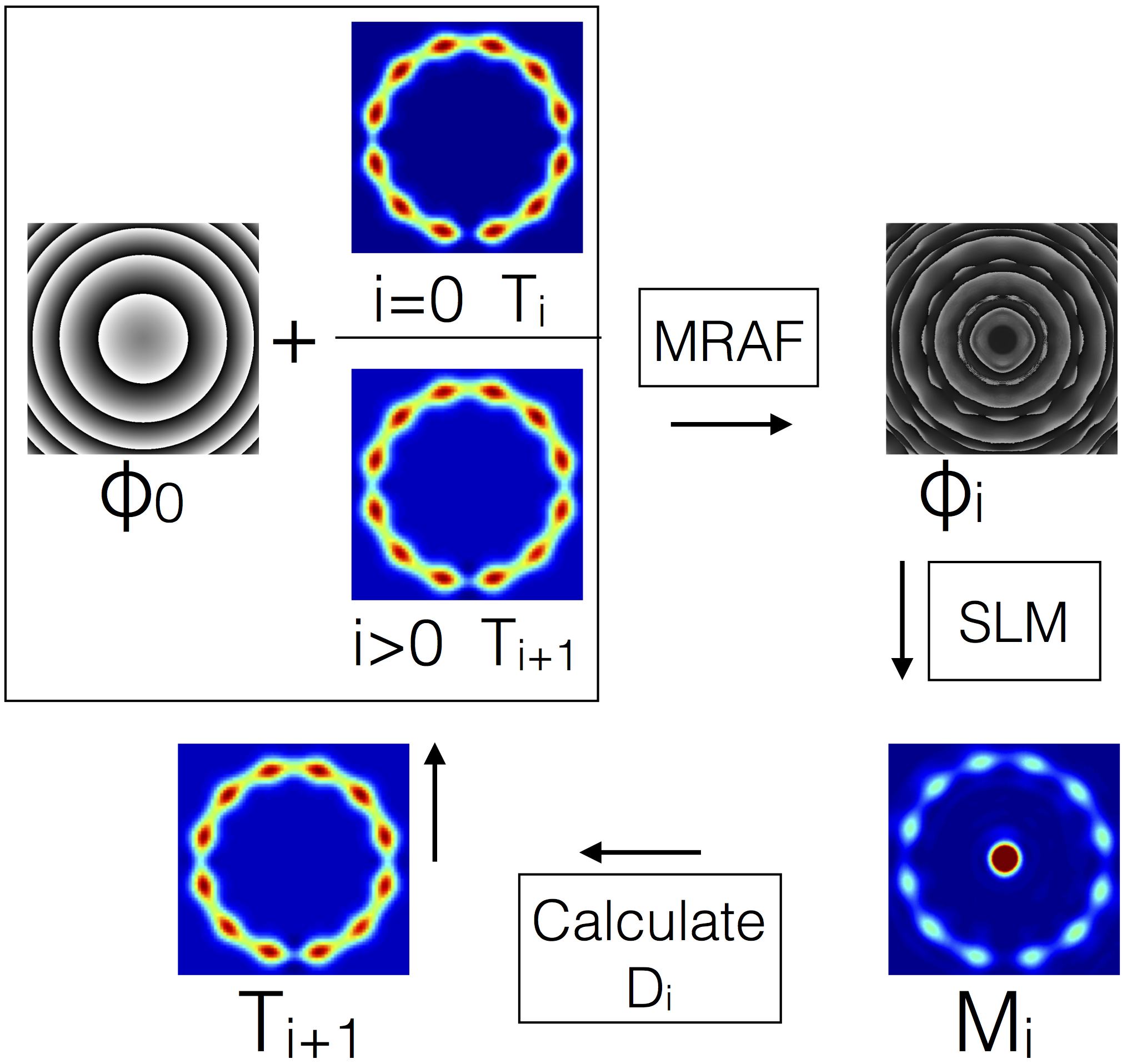}}
    \caption{\label{fig:algor} Our feedback algorithm. Starting at the top left the initial phase and target are used in the MRAF code. This generates the phase guess, $\phi_{i}$, which is uploaded to the SLM and an image captured by the CCD camera, M\textsubscript{i}. This is used to calculate the discrepancy between the image and the original target, and a new target T\textsubscript{i+1} is created. The loop then repeats.}
\end{figure}

\begin{figure}[htb]
    \centerline{\includegraphics[width=0.75\textwidth]{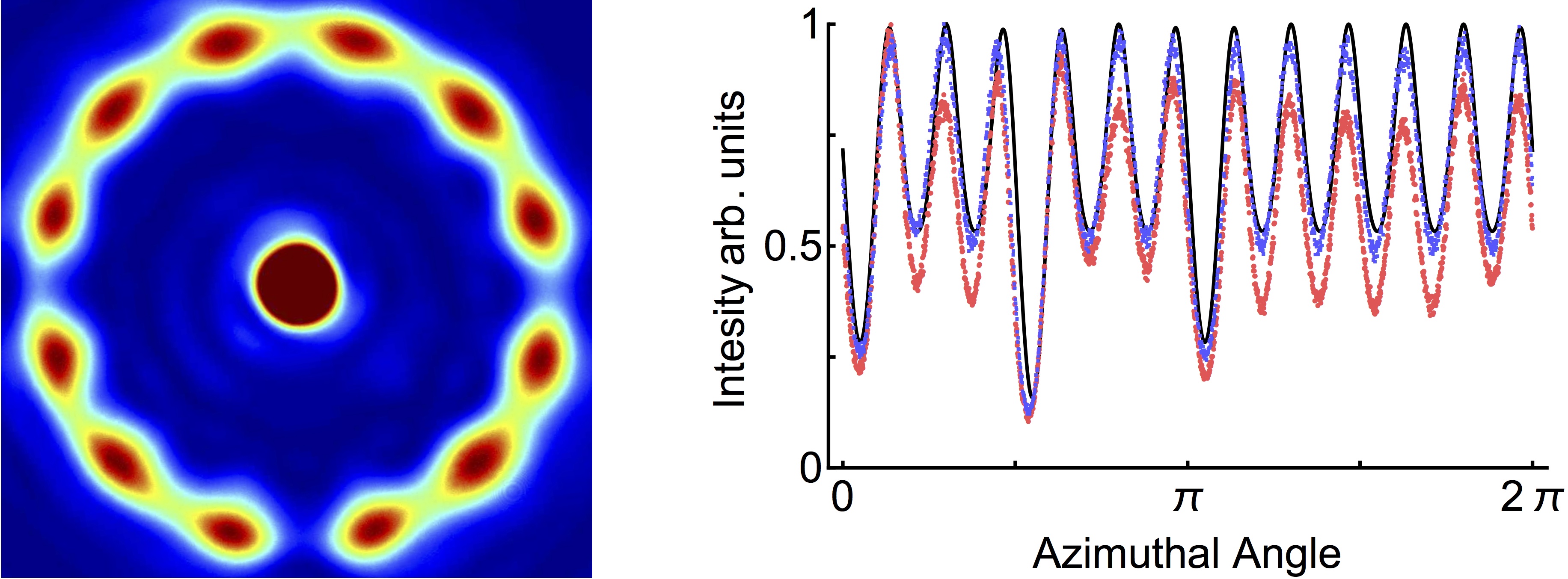}}
    \caption{\label{fig:profile} Left: Final image of the ring lattice after completion of the feedback algorithm. Right: Azimuthal Profile. The solid line plots the target profile.  This is compared to the result after the 1st and 5th iteration of the feedback algorithm (red and blue lines respectively).}
\end{figure}

To increase the accuracy of the output potential we use the computationally generated kinoform and produce an image of the optical potential in the monitoring arm of our system, and use this as a further source of feedback to the MRAF algorithm. Our method is broadly similar to Bruce \textit{et al} \cite{bruce15}, however it is specialised for producing ring-lattices. Fig.~\ref{fig:algor} shows a flow chart of our improved algorithm. In the first step, the target image, T$\textsubscript{i}$, and the initial phase, $\phi_0$, is loaded as an input to the MRAF code. This runs for 20 iterations (this was found to be sufficient to get good convergence in most cases) and outputs a phase kinoform, $\phi$\textsubscript{i}. The kinoform is now applied to the SLM and an image recorded on the camera in the monitoring arm of our system, M\textsubscript{i}. The discrepancy, D\textsubscript{i}, between the original target and the measurement is calculated and used to form an updated target T\textsubscript{i+1}. Here our algorithm differs from \cite{bruce15} as we take the discrepancy to be $D_i=-(M_i^2+T_0^2)/2T_0$. Also, we do not take into account the whole image, the discrepancy is calculated by comparing the maxima and minima around the azimuthal, 1D, profile of the lattice to the target profile. The targets maxima and minima are then adjusted with $T_{i+1}=T_i+\alpha D_i$, where $\alpha$ is a problem specific feedback gain and $i$ the iteration number. The process now repeats with, $\phi_0$ and T\textsubscript{i+1}, as the inputs to the MRAF code. The feedback gain, $\alpha$, is set to be 0.3 to ensure a quick convergence and this process iterates 30 times. At this point the algorithm is complete and the best image from the set M is selected that minimises the discrepancy below 2$\%$. With this method we produce the ring-lattice potential shown in Fig.~\ref{fig:profile} (left), that on the trapping side of our apparatus creates a scaled-down lattice with radius of 5-10 $\mu$m with more than sufficient power to trap ultra-cold atoms. On the right of Fig.~\ref{fig:profile}, the azimuthal profile around the ring lattice is shown. The red curve indicates the profile on the first iteration of the feedback loop. After 5 iterations (blue curve), the algorithm has converged significantly towards the original target (solid line).

\section{Conclusions}
\label{conclu}
We proposed the construction of a flux qubit employing a ring condensate trapped in a regular lattice potential except for three specific lattice points with a reduced tunneling amplitude. The three weak links solution  was originally suggested in quantum electronics to facilitate the function of the system as a qubit. We apply a similar logic leading  to fluxonium from the rf-SQUID:   the continuous quantum fluid, in our system, is replaced by a chain of junctions connecting  the different weak links.  We believe that  the additional lattice helps in adjusting the persistent current flowing through the system. The three weak links architecture, indeed, realizes  a two-level effective dynamics in a considerably enlarged parameter space. 
To this end, we applied a path integral and a exact diagonalization analysis which provide information on the complementary bosonic density regime.  
We comment that, in the rationale adopted in the present paper,  the analytical and numerical work complement each other.  The path integral approach is both of foundational nature and explores the qubit functionality.  Indeed, it provides  essential insights on the physical  reason behind the system's spectroscopy emerging by numerical inspection in  Sec.\ref{qubit}. Moreover, our analytics cover the regime of large fillings and 'medium system size' that is beyond reach of the results based on the exact diagonalization. On the other hand, the latter provides a proof of the effective two level system dynamics, independently on the approximations employed in the analytical treatment (small quantum fluctuations of the bosonic densities) and  for  small system size. Overall, we believe that our system is indeed governed by a two-level dynamics in a wide range of parameters.

With path integral techniques, we obtained the quantum dynamics remaining from the integration of the fast degrees of freedom of the system in the limit of a quantum phase model (valid for large bosonic densities). It turns out that the effective action is similar to  that one  defining the  quantum dynamics  of the superconducting flux qubits \cite{Mooij1999,Feigelman2000}. 
The exact diagonalization study  deal with   systems  of $M\sim 10$ and  low filling($\bar{n} \sim 1-2$) governed by the Bose-Hubbard dynamics (for small bosonic densities). Our analysis demonstrates that the system with three weak links gives more flexibility and freedom for a functional qubit:  contrary to the case of the rf-AQUID, where the only tunable  parameters  are the interaction strength and the height of the barrier of the weak link. Specifically, we demonstrated that interactions can be varied from mild to strong values with $t''/t'$, further parameter to adjust both the qubit energy gap and the quality factor. However, a substantial persistent current is obtained in the limit of mild interactions; in the limit of large interactions, quantum fluctuations of the phase implies a demoted  phase coherence, implying in turn that a  smaller  persistent current can be sustained in the system.

Our exact diagonalization study shows that if one fixes the interaction strength it is possible to adjust both the qubit energy gap and the quality factor by changing $t''$. This provides an extra parameter for defining a good qubit and for realizing single and two-qubit gates (the qubit gates will be addressed elsewhere). Moreover, it is demonstrated that the parameter regime for which the system can perform as a qubit is enlarged, since a functional qubit can be obtained even at moderate $U$ interactions. 
This can be important for the actual realization of the  device since a substantial persistent current is expected to be obtained in this case, Section \ref{qubit}. In Section \ref{EG,QF_N} we have demonstrated  that, compared to the rf-AQUID,  a three-weak links system has more favourable scaling for the energy gap as a function of the filling factor in the limit of mild interactions ($U=1$). Moreover, non-trivial physics emerges due to the overlap between the healing lengths established in the system by the insertion of the weak links. Since there are three healing lengths in the system of a small size their overlap generates a non-trivial density distribution along the ring, which is shown in Section \ref{Densityprofiles}. As noted for the rf-AQUID in \cite{Aghamalyan15}, and also in the present case we observed that  the  interaction and strength of the weak links are responsible for  competing effects (see Fig.\ref{dens}).

In Section \ref{platform} we described our experimental apparatus based on a phase-only spacial light modulator, and our feedback algorithm to produce the ring-lattice potential. Our setup and algorithm can produce ring-lattice potentials with arbitrary link strengths, down to radii of 5-10 $\mu$m with sufficient power to trap Rb\textsuperscript{87} atoms.

\bigskip

{\bf Acknowledgements.-- }
This research has been supported by the National Research Foundation Singapore \& by the Ministry of Education Singapore Academic Research Fund Tier 2 (Grant No. MOE2015-T2-1-101). During this work D.A. was supported by the Agence Nationale de la Recherche (contract no. ANR-12-BS04-0020-01). D.A. would like to thank Andrea Simoni for assisting him with submitting numerical jobs on ED on  the cluster "Simpatix" of Institute for Physical Research.

\appendix


\ \\

\noindent
{\bf References.-- }
\providecommand{\newblock}{}


\bibliographystyle{iopart-num} 

\clearpage
\end{document}